\pdfoutput=1
\documentclass[%
aip,
rsi,%
amsmath,
amssymb,
twocolumn
preprint,%
 reprint,%
]{revtex4-1}
\usepackage{graphicx}
\usepackage{dcolumn}
\usepackage{bm}
\usepackage{hyperref}
\usepackage{lineno}

\hypersetup{
    colorlinks=true,
    linkcolor=blue,
    filecolor=blue,      
    urlcolor=blue,
    citecolor=blue,
}
\usepackage{color,soul}
\usepackage{xspace}
\usepackage{xcolor}
\usepackage[flushleft]{threeparttable}

\usepackage{amsmath}

\newcommand{\degree}{\ensuremath{^{\circ}}\xspace}
\begin{document}
\preprint{}
\title{Interconversion-controlled liquid-liquid phase separation in a molecular chiral model}


\author{Betul Uralcan}
 \affiliation{Department of Chemical and Biological Engineering, Princeton University, Princeton, NJ 08544, USA}
\affiliation{Current address: Department of Chemical Engineering and Polymer Research Center, Bogazici University, Bebek 34342, Istanbul, Turkey}

\author{Thomas J. Longo}
\affiliation{Institute for Physical Science and Technology, University of Maryland, College Park, MD 20742, USA}

\author{Mikhail A. Anisimov}
\affiliation{Institute for Physical Science and Technology, University of Maryland, College Park, MD 20742, USA}
\affiliation{%
Department of Chemical and Biomolecular Engineering, University of Maryland, College Park, MD 20742, USA
}%

\author{Frank H. Stillinger}
\affiliation{Department of Chemistry, Princeton University, Princeton, NJ 08544, USA}
\author{Pablo G. Debenedetti}
\email{pdebene@princeton.edu}
 \affiliation{Department of Chemical and Biological Engineering, Princeton University, Princeton, NJ 08544, USA}

\date{\today}
\newcommand{\sur}[1]{\ensuremath{^{\textrm{#1}}}}
\newcommand{\sous}[1]{\ensuremath{_{\textrm{#1}}}}
\begin{abstract}
Liquid-liquid phase separation of fluids exhibiting interconversion between alternative states has been proposed as an underlying mechanism for fluid polyamorphism, and may be of relevance to protein function and intracellular organization. However, molecular-level insight into the interplay between competing forces that can drive or restrict phase separation in interconverting fluids remains elusive. Here, we utilize {an off-lattice} model of enantiomers with tunable chiral interconversion and interaction properties to elucidate the physics underlying the stabilization and tunability of phase separation in fluids with interconverting states. We show that introducing an imbalance in the intermolecular forces between two enantiomers results in nonequilibrium, arrested phase separation into microdomains. We also find that in the equilibrium case, when all interaction forces are conservative, the growth of the phase domain is restricted only by system size. In this case, we observe phase amplification, in which one of the two alternative phases grows at the expense of the other. These findings provide novel insights on how the interplay between dynamics and thermodynamics defines the equilibrium and steady-state morphologies of phase transitions in fluids with interconverting molecular or supramolecular states.
\end{abstract}

\maketitle

\section{\label{sec:Intro}INTRODUCTION}

Chirality is ubiquitous in nature, and understanding how chiral molecules interact and self-assemble is important to fundamental problems in chemistry, biology, and physics as well as for practical applications. \cite{Blackwell_2003, Cintas_2009, barron, Hoehlig_2015}

Many biological systems are chiral at different levels of organization, including the monomers that constitute proteins, nucleic acids and membranes, as well as the mesoscopic and macroscopic structures that they form, such as the DNA double helix, plant tendril helices and human appendages. \cite{Cintas_2009, barron} From a practical point of view, chirality also plays a key role in many industrial processes.\cite{Sheldrake_1993,Blaser_2013,Blaser_2007} The active ingredients of many drugs are chiral molecules, and their different enantiomers can exhibit significant differences in activity, absorption, selectivity and toxicology. \cite{Lin_2011, Yang_2011, Chhabra_2013}  In addition, engineering the chirality of inorganic materials has recently attracted a great deal of attention in chiral sensing, catalysis, and advanced optical device technologies. \cite{Govan_2019,Govorov_2011,Guerrero_Mart_nez_2011} Molecular level insight on chiral preference and phase behavior is thus desirable to guide synthesis and processing techniques for this rich array of potential technological applications.

While chiral molecules are commonly synthesized as racemic mixtures  (equal proportions of both enantiomers), the molecules of life are comprised of asymmetric building blocks composed of only one of the two possible enantiomers of a given chiral pair (\textit{e.g.}, L-amino acids in proteins, D-deoxyribose in nucleic acids).\cite{Blackwell_2003, Cintas_2009} This is a distinguishing feature of life.  Consequently, chiral symmetry preferences found in nature have long been a subject of investigation\cite{Breslow_2011, Kondepudi_1985, Frank_1953, Soai_1995, Shibata_1996, Lente_2007, Hatch_2010, Ricci_2013, Cintas_2009, Lombardo_2009, Cao_2005, Stillinger_2017}. While many studies have yielded useful insights, the question of how biological homochirality arose in nature, including whether it was predefined or random, remains an area of much current activity. \cite{Mason_1985,Bada_1995, Sandars_2003, Blackmond1, Blackmond2, Blackmond3, Laurente2012741118, Jafarpour_PRL_2015, Jafarpour_PRE_2017} 
Studies of the origin of biological homochirality often involve the search for common principles that seek to explain how a small chiral imbalance can be amplified and subsequently transmitted, giving rise to symmetry breaking.\cite{barron}. In this regard, gaining a fundamental understanding of systems where the chirality of a single molecule influences phase behavior at larger length scales is of significant importance.

{Chiral states are not static in nature, and often the individual molecules of different chiralities may interconvert. Interconversion between alternative molecular states of systems exhibiting phase separation is a ubiquitous phenomenon that has been previously found in many condensed matter systems\cite{R1,R2,R3,R4,R5,R6,R7,R8,R9,GlotPhysRevLett.72.4109, GlotPhysRevLett.74.2034, PLahysRevE.94.022605,P36hysRevLett.75.1674, P37hysRevE.56.3127}.} In this work, we study the interplay between chiral interconversion kinetics and phase behavior in the liquid phase of a three-dimensional, {off-lattice,} flexible tetramer model\cite{Latinwo_2016}. This model consists of chiral tetramer molecules with tunable interconversion rate between {the} two enantiomeric forms. {It also includes a pair potential energy function with a {tunable} chiral bias parameter that {can favor either} locally racemic or heterochiral interactions. {We consider two formulations of the chiral model - one with energy conservation and another one with energy dissipation.}}

{The conserved-force formulation {was introduced recently by Petsev \textit{et al.}\cite{Errata}, but we use different numerical values for the model parameters (see Table I).} In equilibrium, when all interaction forces are balanced, the growth of the phase domain is restricted only by the size of the system. In this case, we observe the phenomenon of phase amplification (``phase bullying''), in which one of the two alternative phases grows at the expense of the other.\cite{Latinwo_2016,Phase_Bullying_2021}} The dissipative formulation considers an imbalance of intermolecular forces resulting from not applying the gradient operator to the chirality-dependent term in the potential energy function.{\cite{Latinwo_2016}} {It corresponds to a nonequilibrium system, in which the imbalance in intermolecular forces facilitates racemization. This racemizing force competes with the equilibrium interconversion and diffusion. At infinite times, this competition leads to the formation of steady-state arrested microphase domains.}

The rest of this paper is organized as follows. In Sec. II, we describe the tetramer model and introduce the model parameters that we use to tune chiral interactions and interconversion kinetics. {In Sec. III.A and {III.B}, we discuss our results on the conserved and dissipative force formulations of the chiral model, respectively.} We show how the interplay between spinodal decomposition and interconversion affects phase separation behavior in the two formulations of the chiral model. In Sec. IV, we provide concluding remarks and suggest some possible directions for future inquiry. {Appendices provide details on the theoretical description of the computational results.}

\section{\label{sec:ModelnMethods}Model and Methods}

This simple chiral tetramer model, inspired by the smallest known chiral molecule in nature, hydrogen peroxide\cite{Ball_2015,Abrahams_1951, Busing_1965}, {was introduced by Latinwo et al.\cite{Latinwo_2016} and subsequently reformulated by introducing an additional 8-body force that produces an energy-conserving force field\cite{Errata}. } A tetramer of the chiral model is composed of 4 monomers along a three-bond backbone (Figure A.1). The instantaneous state of a tetramer is specified by the location of monomers along the backbone at $r_1$, $r_2$, $r_3$ and $r_4$. Specifically, molecules feature left-handed ($A$-type) and right-handed ($B$-type) configurations, and achiral transition states (Figure A.1). The shape of the molecule is determined by the intramolecular potential energy function that includes contributions from bond
stretching, bond angle deformation, and dihedral angle rotation, and is given by

\begin{equation}
\begin{split}
    \Phi^{(1)} (\{\mathbf{r}_i\}) =& \sum_{i=1}^{3} \frac{k_{s}}{2} \left(r_{i,i+1} - b\right)^{2} \\& + \sum_{i=1}^{2} \frac{k_{b}}{2} \left(\Theta _{i} - \frac{\pi}{2} \right)^{2} + k{_{d}} \textup{cos}^{2} \phi 
\end{split}
\end{equation}
where $r_{i,i+1}$ is the instantaneous distance between sites $i$ and $i+1$, $b$ represents the equilibrium bond length, {\it{$\Theta_i$}} and {\it{$\phi$}} are the bond and dihedral angles, and {\it{$k_s$}}, {\it{$k_b$}}, and {\it{$k_d$}} are the force constants for bond stretching, angle bending, and dihedral motion, respectively. The dihedral force constant {\it{$k_d$}} controls the rigidity of the dihedral angle of a tetramer and determines the rate of interconversion between a pair of mirror image configurations. {The chiral model parameters are given in Table I. The model's behavior is only a function of dimensionless (reduced) variables. The parameters in Table I are suggested physical constants for translating reduced units (e.g., $P^* = P\sigma_{tt}^{3}/\epsilon_0$) into actual physical quantities (e.g., $P = P^*\epsilon_0/\sigma_{tt}^{3}$).}

\begin{table}
\begin{threeparttable}
\centering
\caption{Parameters for the chiral model\textsuperscript{a}} 
\begin{tabular}{|c|c|c|c|c|c|c|c|}
\hline
 & $k_s$              & $k_b$                      & $k_d$                 & $b$                  \\ \hline
\multicolumn{1}{|c|}{Actual}    & \multicolumn{1}{c|}{196 } & \multicolumn{1}{c|}{193 } & \multicolumn{1}{c|}{0.003-7.76 } & \multicolumn{1}{c|}{3.7 } \\
 & kcal/mol {\AA}$^2$              & kcal/mol                      & kcal/mol                 & \AA                  \\  \hline
\multicolumn{1}{|c|}{Reduced} & \multicolumn{1}{c|}{8003} & \multicolumn{1}{c|}{643.7} & \multicolumn{1}{c|}{0.001-25.86} & \multicolumn{1}{c|}{1.0583} \\ \hline
\end{tabular}
\begin{tablenotes}
  \item \textsuperscript{a} Throughout the paper, distances are expressed in units of $\sigma_{tt}=$ 3.5 \AA (see Eq. 3), energies in units of $\epsilon_0 = $ 0.3 kcal/mol (see Eq. 4), temperatures in units of $\epsilon_0/k$ = 150.9 K ($k$ is Boltzmann's constant), pressures in units of $\epsilon_0\sigma_{tt}^{-3}$ = 486 bar, densities in units of $\sigma_{tt}^{-3}$ = 0.023 \AA$^{-3}$, and time in units of $\sigma_{tt}\sqrt{m^*/\epsilon_0}$ = 0.3124 ps, where $m^*$ = 1 g/mol. The mass of a monomer was set to 8.5$m^*$, yielding a molecular weight of 34 g/mol.
\end{tablenotes}
  \end{threeparttable}
\label{Table1}
\end{table}

In order to monitor the chirality and control the intermolecular interactions between tetramer pairs, we define a scalar chirality measure, $-1\leq \zeta \leq 1$
\begin{equation}
\zeta (\mathrm{\mathbf{r}}_1, \mathrm{\mathbf{r}}_2, \mathrm{\mathbf{r}}_3, \mathrm{\mathbf{r}}_4) = \frac{\mathrm{\mathbf{r}}_{12} . (\mathrm{\mathbf{r}}_{23} \, \times \, \mathrm{\mathbf{r}}_{34})}{|\mathrm{\mathbf{r}}_{12}||\mathrm{\mathbf{r}}_{23}||\mathrm{\mathbf{r}}_{34}|}
\end{equation}
where, for each tetramer, the chirality of the enantiomer is determined by the sign of $\zeta$. This measure attains its lower and upper limits for the mirror image left-handed ({\it{A}}-type) and right-handed ({\it{B}}-type) configurations and goes to zero for the achiral transition states illustrated in Figure A.1 (see Appendix A). 

The model's intermolecular potential energy function between two molecules $\alpha$ and $\gamma$ is given by 
\begin{equation}\label{Eq_Phi2}
\Phi ^{(2)} (\left \{ \mathbf{r}_{i}^{\alpha} \right \}, \left \{ \mathbf{r}_{j}^{\gamma } \right \}) = \sum_{i,j=1}^{4} \epsilon_{tt} (\zeta ^{\alpha},\zeta ^{\gamma})\upsilon \left (\frac{\left |\mathbf{r}_{i}^{\alpha}-\mathbf{r}_{j}^{\gamma}  \right |}{\sigma_{tt}} \right )
\end{equation}
where the double sum runs over each site on each tetramer, ${\sigma_{tt}}$ is the pair potential distance parameter which specifies the range of the site-site interactions, $\upsilon$ is the 12-6 Lennard-Jones function, and $\epsilon_{tt}$ is the strength of the interaction energy between tetramers. The interaction energy term $\epsilon_{tt}$ can be tuned to favor/disfavor homochiral/heterochiral interactions between tetramers $\alpha$ and $\gamma$, and is given by 
\begin{equation}\label{Eq_Epstt}
\epsilon_{tt} (\zeta ^{\alpha},\zeta ^{\gamma}) = \epsilon_{0} \left [ 1 + \lambda \zeta(\mathbf{r}_{i=1,2,3,4}^{\alpha})\zeta(\mathbf{r}_{i=1,2,3,4}^{\gamma}) \right ]
\end{equation}
where $\lambda$ is the chirality renormalization parameter such that $\lambda$ $<$ 0 favors heterochiral interactions between tetramers $\alpha$ and $\gamma$ (i.e., lower energy when $\alpha$ and $\gamma$ have opposite chiralities), $\lambda$ $>$ 0 favors homochiral interactions between tetramers $\alpha$ and $\gamma$ (i.e., lower energy when $\alpha$ and $\gamma$ have the same chirality), and $\lambda$ = 0 represents a bias-free scenario\cite{Latinwo_2016}. We note that while its numerical value is an adjustable model parameter, the chirality renormalization parameter effectively represents the local binding preferences of real chiral molecules. For instance, aspartic acid and glutamic acid both display homochiral bias ($\lambda>0$) evident from their enantiopure crystal structures\cite{viedma2001enantiomeric}, while serine and histidine both exhibit heterochiral bias ($\lambda<0$) evident from their racemic crystal structures\cite{histidine}. 

{The two formulations of the chiral model differ in the forces that stem from the intermolecular potential defined in Equation 3. In the first formulation\cite{Latinwo_2016}, only the Lennard-Jones force on monomers with interaction energy $\epsilon_{tt}$ is considered. However, the gradient operator that yields the forces corresponding to the pair potential defined in Equation 3 applies rigorously to the entire argument inside the double summation. Hence this model introduces dissipation through a non-conservative force. In the second formulation\cite{Errata}, the gradient properly operates on the Lennard-Jones and on the chirality-dependent energy pre-factor, producing a coarse-grained energy-conserving force field. That both models are coarse-grained follows from the fact that the interaction energy between two tetramers depends explicitly on their respective chiralities, a tetramer-level (as opposed to site-level) quantity.}

\section{\label{sec:Results}Results and Discussion} 
Computational results for the two formulations of the chiral model, with conservative and dissipative intermolecular forces, are presented in this section. The time evolution of both formulations is described based on a generalized Cahn-Hilliard theoretical model of phase separation which includes molecular interconversion of species. The model is a mean field approach, which is accurate away from the critical point.

\subsection{Chiral Model with Conservative Intermolecular Forces}
\begin{figure}[h!]
    \centering
    \includegraphics[width=\linewidth]{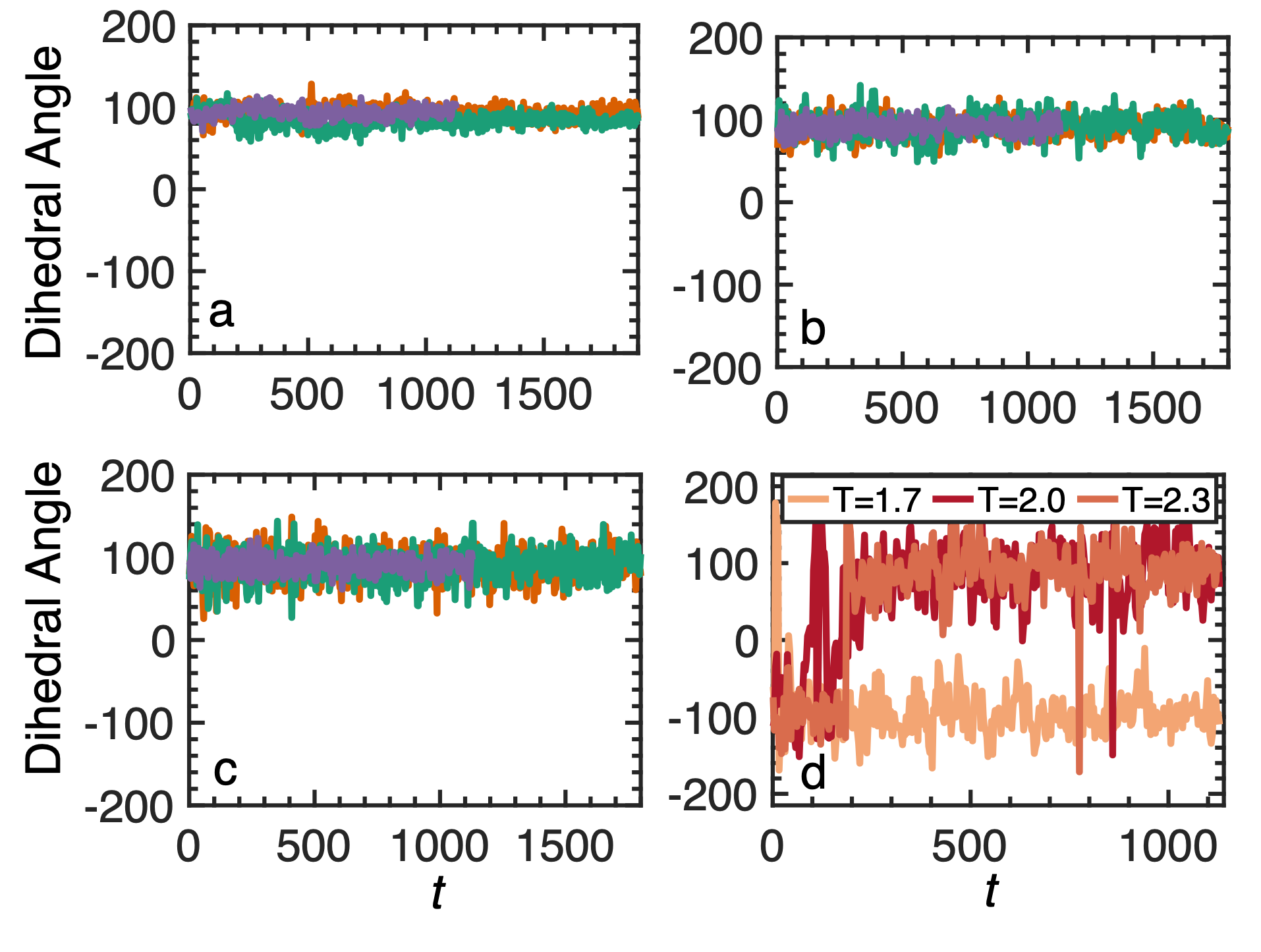}
    \caption{ {Time dependence of the instantaneous dihedral angle of a typical tetramer in a racemic mixture at $P=0.1$ for the conservative-force formulation of the model, at several values of $T$ and $k_d$. a) $T = 0.6$, b) $T = 1.7$, and c) $T = 2.3$, with $k_{d}$ = 5 (green), $k_{d}$ = 9.86 (orange) and  $k_{d}$ = 19.86 (purple). d) Behavior of the dihedral angle at a very low value of the dihedral constant, $k_{d}$ = 0.001.} }
    \label{fig:Interconversion_8Force}
\end{figure}

In Figure \ref*{fig:Interconversion_8Force}, the time dependence of the dihedral angle of a single tetramer in a racemic mixture of 1000 tetramers, at different temperatures and for a range of values of the dihedral force constant, is shown for the chiral model with conservative forces. It can be seen that the enantiomers do not interconvert within the simulation times sampled here (Figure \ref*{fig:Interconversion_8Force}a-c). Interconversion between the two enantiomorphs is achieved only at much lower $k_d$ values ($k_d$=0.001 in Figure \ref*{fig:Interconversion_8Force}d).

When a binary mixture with equal concentrations of interconverting molecules is quenched from a high temperature to a low temperature, below the critical temperature of demixing, there are two processes that may occur. The system may either phase separate, through a process known as spinodal decomposition\cite{Cahn_1965}, or one of the two alternative phases will grow at the expense of {the other}, a process known as phase amplification\cite{Phase_Bullying_2021}. {The chiral model with conservative intermolecular forces always undergoes phase amplification below the critical demixing temperature. This is because in a system where molecules can interconvert, the number of molecules of each type (chirally distinct enantiomers in the present case) is not a conserved quantity, and hence the system minimizes its free energy by avoiding the energetic penalty associated with the formation of an interface, and one of the two alternative phases grows at the expense of the other. Which of the two phases grows is of course a stochastic event.}

{The above considerations apply strictly only at true thermodynamic equilibrium. From a numerical point of view, it is important to understand that the ``stiffness'' of the force constant for the dihedral angle, $k_d$, determines the ease with which such equilibrium can be attained. The mean frequency with which an individual molecule is able to switch its chirality varies in the opposite direction to any $k_d$ variation. Thus,  below the critical temperature for demixing, and for small enough values of $k_d$, interconversion occurs frequently and the system is able to attain true equilibrium, resulting in phase amplification. On the other hand, for sufficiently large values of $k_d$, interconversion is increasingly rare, the system is under diffusive control, and phase separation, rather than amplification, occurs on practical time scales accessible to simulation, even if the system is under the action of conservative forces.}

The mixed diffusion and interconversion dynamics of the chiral model can be described through a Cahn-Hilliard theoretical model of phase separation which includes interconversion between species\cite{Phase_Bullying_2021}. According to this model, the dynamics of such a system is characterized by a growth rate (see Appendix B.1 and reference\cite*{MFT_PT_2021}) of the form
\begin{equation}\label{Eq-Amplification_Factor}
    \omega(q) = - \Delta T (LR_0^2 + Mq^2)(1-\xi^2 q^2)
\end{equation}
where $q$ is the wave number (=2$\pi$/$r$), $R_0$ is the size of the tetramer (adopted here as $R_0=1$ in reduced units), $\Delta T = T/T_\text{c} - 1$ is the distance to the critical temperature of demixing (negative when the system is in the unstable region), and $M$ and $L$ are the diffusion and interconversion Onsager kinetic coefficients respectively. {$\xi^2$ is the square of the correlation length of mesoscopic concentration fluctuations, which diverge at the spinodal as $\xi^2 \sim 1/(-\Delta T)$, in the mean field approximation}. The self-diffusion (mobility) coefficient is given by $M \approx k T/6\pi\eta R_0$ (where $\eta$ is the shear viscosity). The interconversion Onsager kinetic coefficient, $L$ depends on the strength of the rigidity spring constant, $k_d$, such that it becomes zero in the limit when $k_d\to\infty$ and diverges when $k_d\to 0$. Therefore, for large values of $k_d$ ($M\gg L$), phase separation is expected, while for very small values of $k_d$ ($L\gg M$), phase amplification is predicted. 

\begin{figure}[h]
\includegraphics[width=\linewidth]{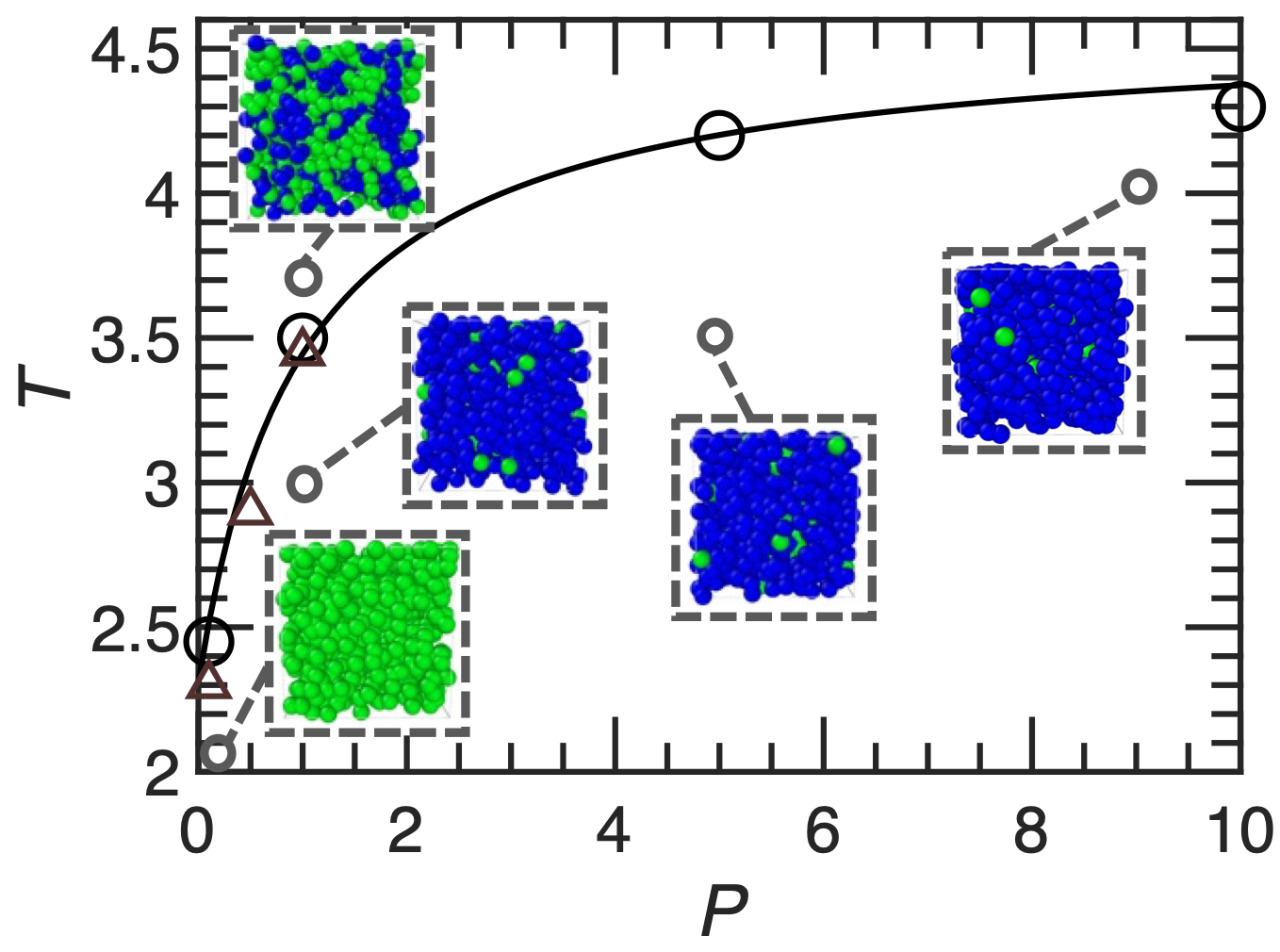}
\caption{The phase diagram showing chiral phase amplification for the chiral model with conservative intermolecular forces, heterochiral bias parameter $\lambda = 0.5$, and rigidity spring constant $k_d = 0.001$. The circles on the solid curve are the computational data for the critical temperature of equilibrium phase separation and the curve is the fit of Equation~(\ref*{Eq-TCPressure}). The images show snapshots of the equilibrium states for the pressures $P=0.1$, $P=1$, $P=5$, and $P=10$ below the critical temperature and at $P=1$ above the critical temperature. The triangles show the prediction of the critical temperature from the extrapolation of the chiral model with dissipative intermolecular forces to the limit $k_d\to\infty$ (see Sec. III.B).}
\label{fgr:FigureTen}
\end{figure}

Figure \ref*{fgr:FigureTen} shows the phase diagram of the chiral model in the case when $k_d = 0.001$ ($L\gg M$) and interconversion dynamics controls the phase behavior of the system. The pressure dependence of the critical temperature is empirically described by
\begin{equation}\label{Eq-TCPressure}
    T_\text{c}(P) = T_\text{c}(P=0) +  \frac{\alpha P}{1 + P}
\end{equation}
{where $T_\text{c}(P=0) = 2.19$ and $\alpha = 2.43$}. {We point out that although $k_d$ primarily determines the interconversion rate, it also affects the system's equilibrium thermodynamics. For example, the total pair interaction energy shows a small but non-zero dependence on $k_d$ at otherwise identical thermodynamic conditions (Appendix {C}). Accordingly, one expects the critical temperature to depend on $k_d$. This effect is at the limit of detectability, and for numerical purposes, $k_d$ can be considered as controlling interconversion kinetics while having at most a modest effect on the system's thermodynamics.}

As shown in Figure \ref*{fgr:FigureTen}, where $\lambda = 0.5$, above the critical temperature, a homogeneous mixture of A- and B- enantiomers is observed throughout the simulation box. The apparent mesoscopic inhomogenities shown by the snapshot in Figure \ref*{fgr:FigureTen} above the critical temperature are attributed to the growing correlation length of concentration fluctuations in the critical region. Below $T_\text{c}$, phase amplification, in which one phase grows at the expense of the other, occurs. Note that by quenching the racemic mixture below $T_\text{c}$, due to the effect of phase amplification\cite{Phase_Bullying_2021}, the system equilibrates arbitrarily to either A- or B-type enriched enantiomer phases as illustrated in Figure \ref*{fgr:FigureTen}, thereby establishing a chiral preference. 

\subsection{Chiral Model with Dissipative Intermolecular Forces}
Dissipative intermolecular forces prevents the chiral model from relaxing to an equilibrium state; instead, this system evolves into a nonequilibrium steady state. This effect can be accounted for by a modification in the growth rate formula, Equation \ref*{Eq-Amplification_Factor}, where the energy dissipation causes {forced} racemization of species. This {forced} racemization competes with the inherent equilibrium interconversion. 

\begin{figure}[h]
 \includegraphics[width=3.25in]{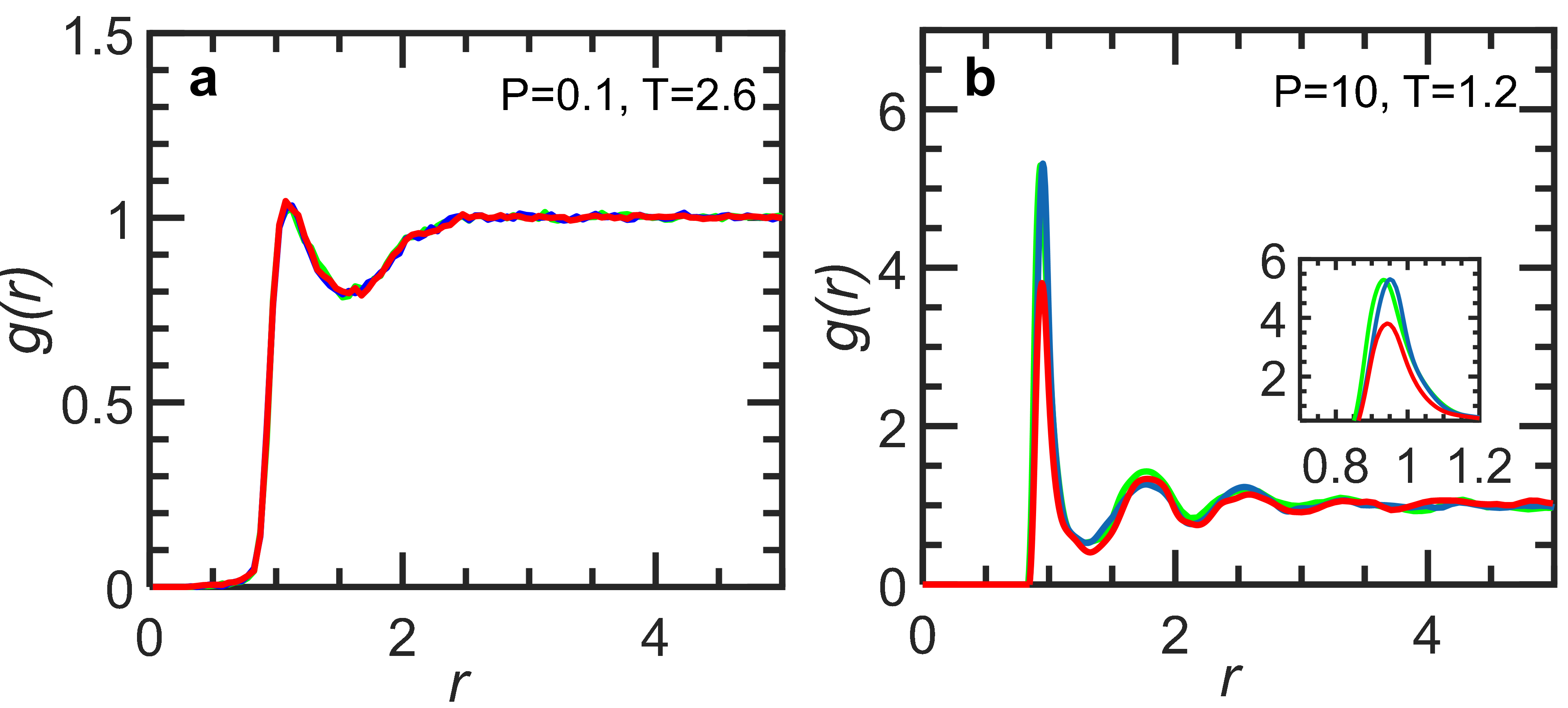}
\caption{Pairwise local structure in the absence of an explicit energetic bias at $k_d$=25.86 for the dissipative-force formulation of the chiral model. Tetramer-tetramer center of mass radial distribution functions for {\it{A-A}} (green), {\it{B-B}} (blue), {\it{A-B}} (red) pairs indicate a) mixing at low pressure ($P$=0.1) and high temperature ($T$=2.6) and, b) local homochiral bias at high pressure ($P$=10) and low temperature ($T$=1.2). The inset shows the overlap between the green and blue curves. 
}
\label{fgr:FigureOne}
\end{figure}

We first studied the phase behavior of the dissipative chiral model in the absence of an explicit energetic bias for homochiral interactions ($\lambda =$ 0). Figure \ref*{fgr:FigureOne} shows the pairwise tetramer-tetramer center
of mass radial distribution profiles $g(r)$ for a bias-free system of 1000 tetramers at $t$=10$^6$ after starting from a homogeneous racemic configuration. While complete phase separation is not observed on the time scale of the simulations, the partial radial distribution profiles illustrate the spontaneous enhancement of homochiral interactions in the tetramer model at $P$=10 and $T$=1.2, illustrated by the sharp first peaks of $g(r)$ profiles for {\it{A-A}} and {\it{B-B}} tetramer pairs. Nevertheless, without an explicit bias parameter, enhancement of homochiral interactions is observed only at very high pressures and low temperatures, where diffusion and chiral interconversion kinetics are slow. Consequently, in this study, we employ a homochiral bias parameter that energetically favors homochiral interactions ($\lambda$ = 0.5) to study the liquid-liquid phase separation of the chiral system.  

\begin{figure}[h]
 \includegraphics[width=3.25in]{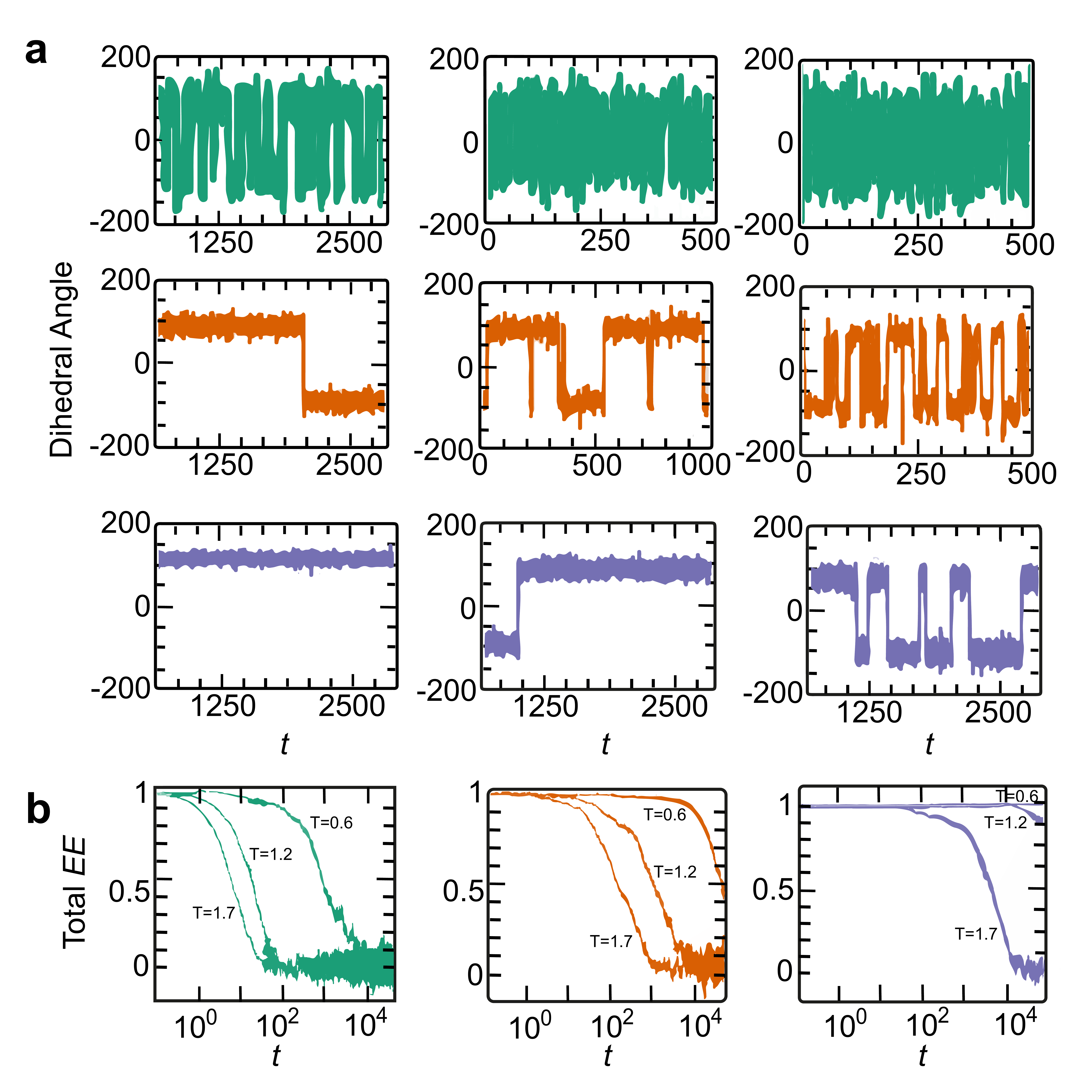}
\caption{Interconversion and racemization kinetics of the dissipative-force formulation of the chiral model. Temporal evolution of a) a single tetramer in a racemic mixture, with dihedral force constant (from top to bottom) {\it{$k_d$}} = 5 (green), {\it{$k_d$}} = 9.86 (orange) and {\it{$k_d$}} = 19.86 (purple) at {$P$ = 0.1}, T (left to right) = 0.6, 1.2 and 1.7, and b) the total enantiomeric excess (see Equation \ref*{Eq_TotEE}) of an initially enantiopure system, with dihedral force constant (from left to right) {\it{$k_d$}}=5 (green), {\it{$k_d$}}=9.86 (orange) and {\it{$k_d$}}=19.86 (purple) at {$P$ = 0.1}, $T$ = 0.6, 1.2 and 1.7.}
\label{fgr:FigureTwo}
\end{figure}

Figure \ref*{fgr:FigureTwo}a depicts the strong {\it{$k_d$}} dependence of chiral interconversion kinetics by showing the temporal evolution of the dihedral angle of a single tetramer in a racemic mixture of 1000 tetramers. At high {\it{$k_d$}} and low $T$, the tetramer persists longer in the vicinity of the stable enantiomorphs ($\phi$ $\approx$ -90\degree for {\it{A}}-type and $\phi$ $\approx$ 90$\degree$ for {\it{B}}-type tetramers). In particular at {\it{$k_d$}}=19.86 and $T=$0.6, it reaches the limit at which no interconversion is observed within the simulation time ({\it{$\tau_{obs}$}}). In this limit of slow interconversion, where the characteristic interconversion time of a tetramer {\it{$\tau_{INC}$}}, defined as the average time required for a tetramer to switch chirality, is much longer than the total simulation time {\it{$\tau_{obs}$}}, the system behaves thermodynamically as a binary mixture of enantiomers that do not interconvert.  In contrast, at low {\it{$k_d$}} and high $T$, the tetramer interconverts between its two stable enantiomorphs very rapidly. At these conditions, the achiral transition states ($\phi$ $\sim$ -180\degree, 0\degree, 180$\degree$ for {\it{cis}} and {\it{trans}} configurations, respectively) also become more accessible. In this opposite limit of very fast interconversion, the system can be treated thermodynamically as a single-component fluid, since the characteristic interconversion time {\it{$\tau_{INC}$}} is much shorter than the total simulation time {\it{$\tau_{obs}$}}.

{Comparing the interconversion kinetics of the dissipative force formulation of the chiral model (Figure \ref*{fgr:FigureTwo}a) with the conservative force formulation (Figure \ref*{fig:Interconversion_8Force}) one can see that dissipation lowers the barrier to enantiomer racemization, thereby causing the chiral interconversion rate to increase. The increase in the interconversion kinetics in the dissipative formulation {reflects} the absence of the additional force contribution arising, in the conservative case, from applying the spatial gradient operator to the chirality-dependent characteristic energy $\epsilon_{tt}$.}

Next, to investigate the effect of the interconversion rate on the racemization kinetics, we study the time-dependent behavior of the average chirality (the total enantiomeric excess) of the tetramer system, starting from an enantiopure configuration (Figure \ref*{fgr:FigureTwo}b). The total enantiomeric excess ({\it{EE}}) for the bulk tetramer system is defined by
\begin{equation}\label{Eq_TotEE}
    \textrm{Total} \: EE=\frac{{N_{A}-N_{B}}}{N_{A}+N_{B}} 
\end{equation}
where $\it{N_A}$  and $\it{N_B}$ are the number of {\it{A}}- and {\it{B}}-type tetramers, respectively. At low {\it{$k_d$}} and high $T$, systems that start from enantiopure configurations (total {\it{EE}} = 1) tend rapidly towards racemic mixtures with vanishing average chirality, consistent with the corresponding tetramer interconversion kinetics profiles. Conversely, the total {\it{EE}} of high {\it{$k_d$}} and low $T$ systems remains constant within {\it{$\tau_{obs}$}}, demonstrating that these systems can be treated as non-interconverting binary mixtures (Figure 4b, {\it{$k_d$}} = 19.86, $T$ = 0.6 and 1.2). 

\begin{figure}[h!]
    \centering
    \includegraphics[width=3.25in]{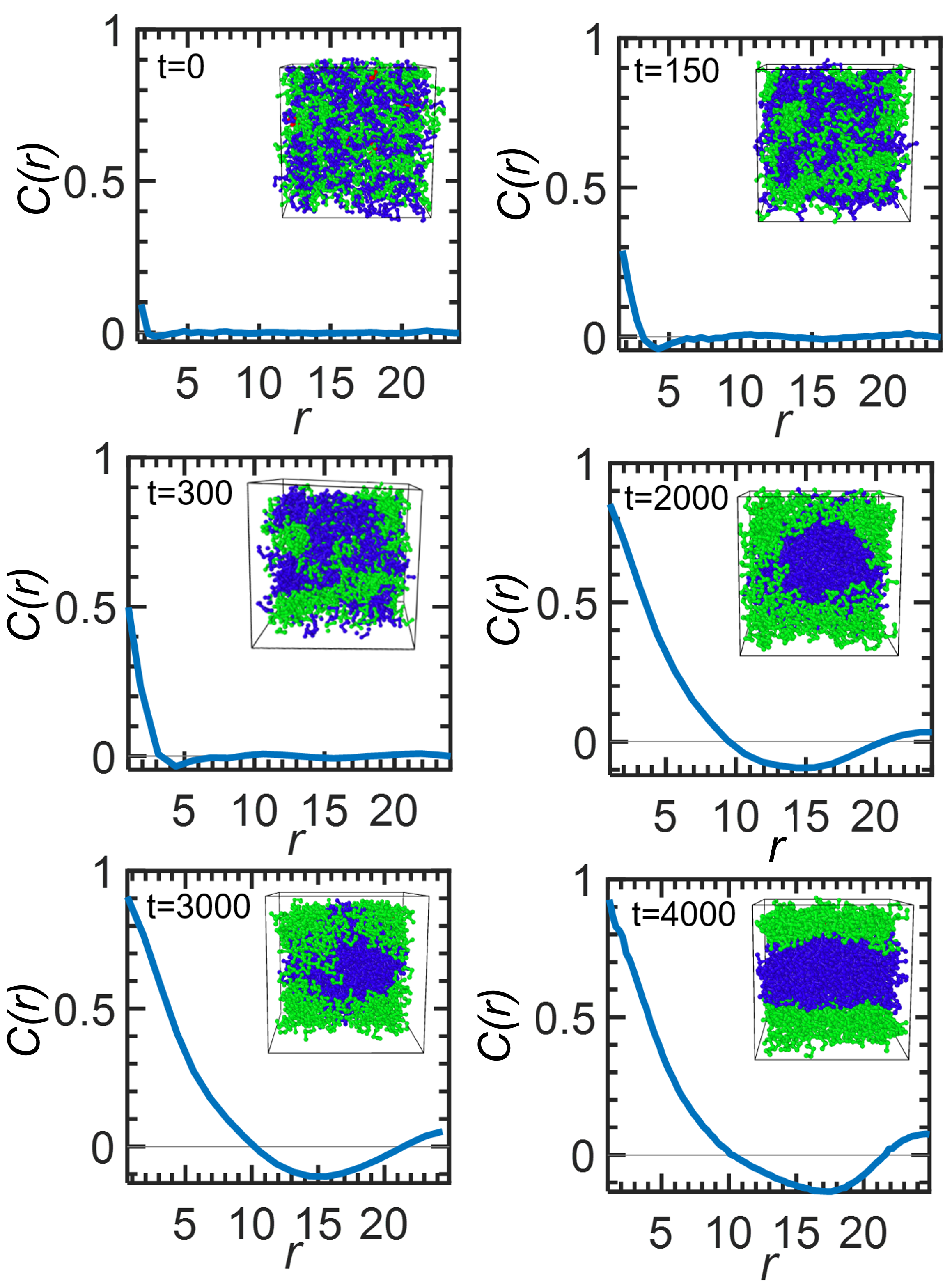}
    \caption{Phase separation kinetics of the dissipative-force formulation of the chiral model. Temporal evolution of phase separation depicted by the instantaneous spatial correlation profiles and representative snapshots for a racemic mixture ({\it{$k_d$}}=9.86, $T=0.8$, and  $P$=0.1).}
    \label{Fig_Figure3a}
\end{figure}

In order to understand the growth of the spatial correlation of enantiomorphs as a function of time, we compute a time- and position- dependent order parameter $ee (r, t)$ given by 
\begin{equation}
ee (r, t) = \frac{N_{A}(r,t)-N_{B}(r,t)}{N_{A}(r,t)+N_{B}(r,t)} 
\end{equation}

{Equation 8 applies to the case where the molecule at $r = 0$ is an A enantiomorph. If instead a B molecule is at $r = 0$, the numerator changes to $N_B(r, t) - N_A(r, t)$.} This order parameter is inspired by Cahn-Hilliard theory\cite{CahnHillard} where $ee (r, t)$ defines the length scale dependent ordering process in a {phase-separating} binary mixture when quenched below the coexistence {and spinodal lines}. The physical significance of the order parameter is as follows: the order parameter approaches unity when the neighbors of an enantiomorph at a distance $r$ are of the same type as the molecule at the origin, signifying compositional inhomogeneity (and possibly phase separation) at the length scale {\it{r}}. When the enantiomorphs are homogenously mixed, $ee (r, t)$ decays to zero. We compute the spatial correlation of $ee (r, t)$ using
\begin{equation}
C(r,t)=\frac{\left \langle  ee(r',t)ee(r'+r,t)\right \rangle-\left \langle ee(r',t)\right \rangle \left \langle ee(r'+r,t)\right \rangle}{\left \langle ee(r',t)ee(r',t)\right \rangle-\left \langle ee(r',t)\right \rangle \left \langle ee(r'+r,t)\right \rangle}
\end{equation} 

Figure \ref*{Fig_Figure3a} shows the temporal evolution of the $C(r, t)$ profiles for a homogeneously mixed racemic mixture of 1000 particles with {\it{$k_d$}} = 11.86, quenched from $T$ = 2.6 to $T$ = 0.8. The distance {\it{R}} at which $C (r, t)$ {first} decays to zero ({\it{r=R}}) gives the average domain size, hence the characteristic length scale of the phase separation. The two enantiomorphs {\it{A}} and {\it{B}} are marked as green and blue tetramers, respectively. Initially, the spatial correlation function $C(r, t)$ fluctuates around zero, signifying that the enantiomorphs are homogeneously mixed before the quench, {\it{R}}({\it{t}} = 0) = 0. After the quench, the tetramers interconvert and diffuse in such a way that local enantiopure configurations start to be favored. At early times ({\it{t}} = 150), the correlation length {\it{R}} is small, indicating the formation of small enantiomorph clusters. At later stages ({\it{t}} = 300), phase separation proceeds rapidly as these small domains merge. The rapid phase separation process slows down after {\it{t}} = 300, and coarsening of the {\it{A}}- and {\it{B}}-rich phases takes over. This final stage completes when the rough edges of the interface between the two enantiomorph phases are smoothed ({\it{t}} = 4000).

\begin{figure}[h]
\includegraphics[width=3.25in]{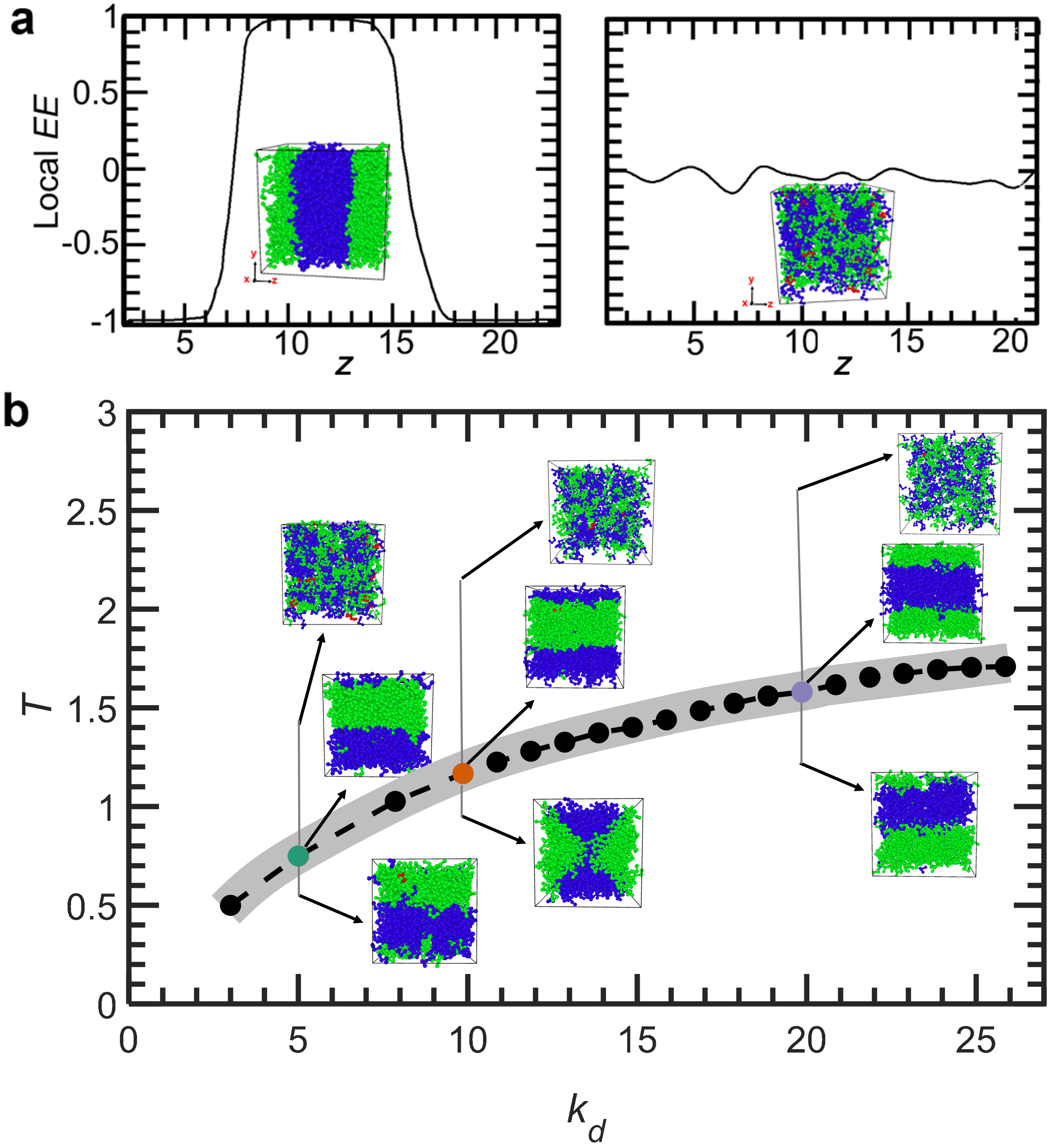}
\caption{Chirality-induced liquid-liquid phase separation with explicit bias that favors homochiral interactions ($\lambda$ = 0.5) in the dissipative-force formulation. a) The local enantiomeric excess profiles as a function of distance normal to the planar interface $z=0$ for a fully phase separated system ($\it{R_{\infty}=R_{max}}$) (left) and a system with small enantiomorph clusters ($\it{R_{\infty}<R_{max}}$) (right). b) Onset temperature, $T^*$, for liquid-liquid phase separation at the length scale of the simulation box as a function of dihedral angle force constant, $k_d$, at $P=0.1$. The colored points represent the conditions for the snapshots of the chiral system above, at, and below $T^*$. The dashed line is given as a guide for $T^*$. {The grey lane illustrates the uncertainty in the definition of $T^*\approx \pm 0.1$.}
}
\label{fgr:FigureSix}
\end{figure}

{In Figure \ref*{fgr:FigureSix}, we illustrate kinetically arrested liquid-liquid phase separation for the dissipative-force formulation of the chiral model at the scale of the simulation system (wave number, $q = 2\pi/\ell$, where $\ell$ is the length of the simulation box). In particular, starting from a homogeneous racemic mixture of enantiomers, we study the steady-state behavior of the tetramer system upon quenching to a temperature and pressure of interest. Figure \ref*{fgr:FigureSix}a shows representative local enantiomeric excess profiles for a phase-separated ($k_d$=19.86 and $T$=1) and a homogeneously mixed system ($k_d$=19.86 and $T$=2) that have reached steady-state.} {Figure \ref*{fgr:FigureSix}b shows the apparent onset temperature of liquid-liquid phase separation ($T^{*}$) defined as the temperature where the growing steady-state domain size, $R_\infty$, reaches the size of the simulation box, $R_{\text{max}}\sim\ell$. This temperature increases with the rigidity of the dihedral angle of the tetramers. Phase separation can be considered fully developed below the onset temperature. However, interconversion frustrates complete phase separation above the onset temperature. As illustrated by the simulation snapshots of Figure \ref*{fgr:FigureSix}b, above $T^*$ we observe phase separated domains of smaller sizes $\it{R_{\infty}}<\it{R_{max}}$.}

\begin{figure}[h!]
\includegraphics[width=\linewidth]{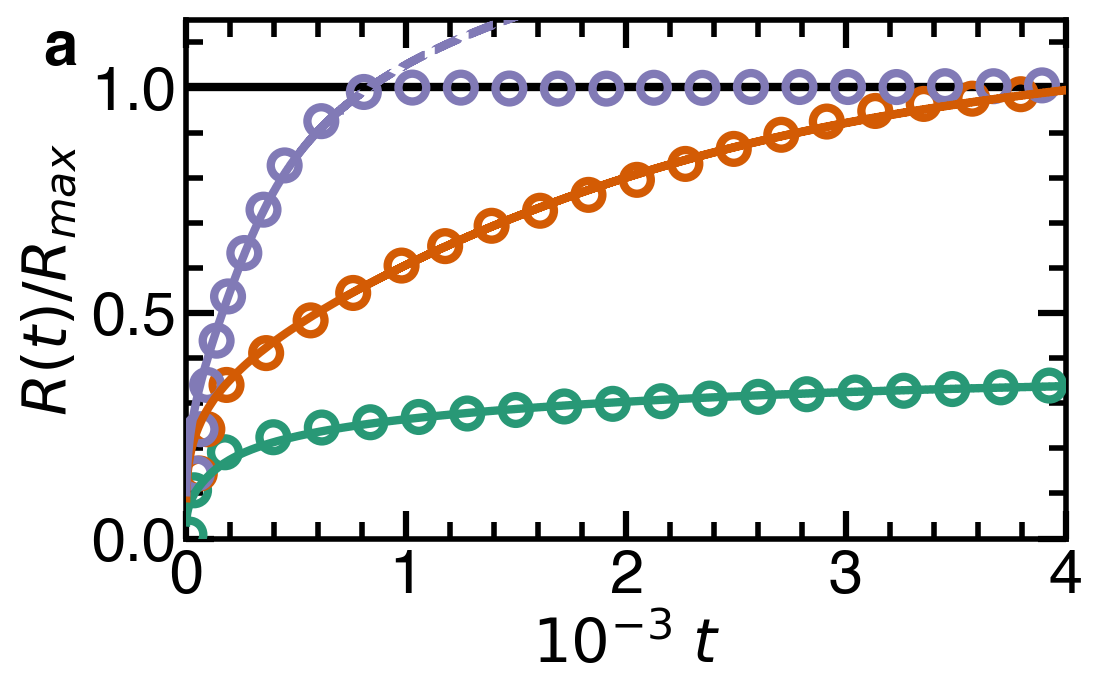}
\includegraphics[width=\linewidth]{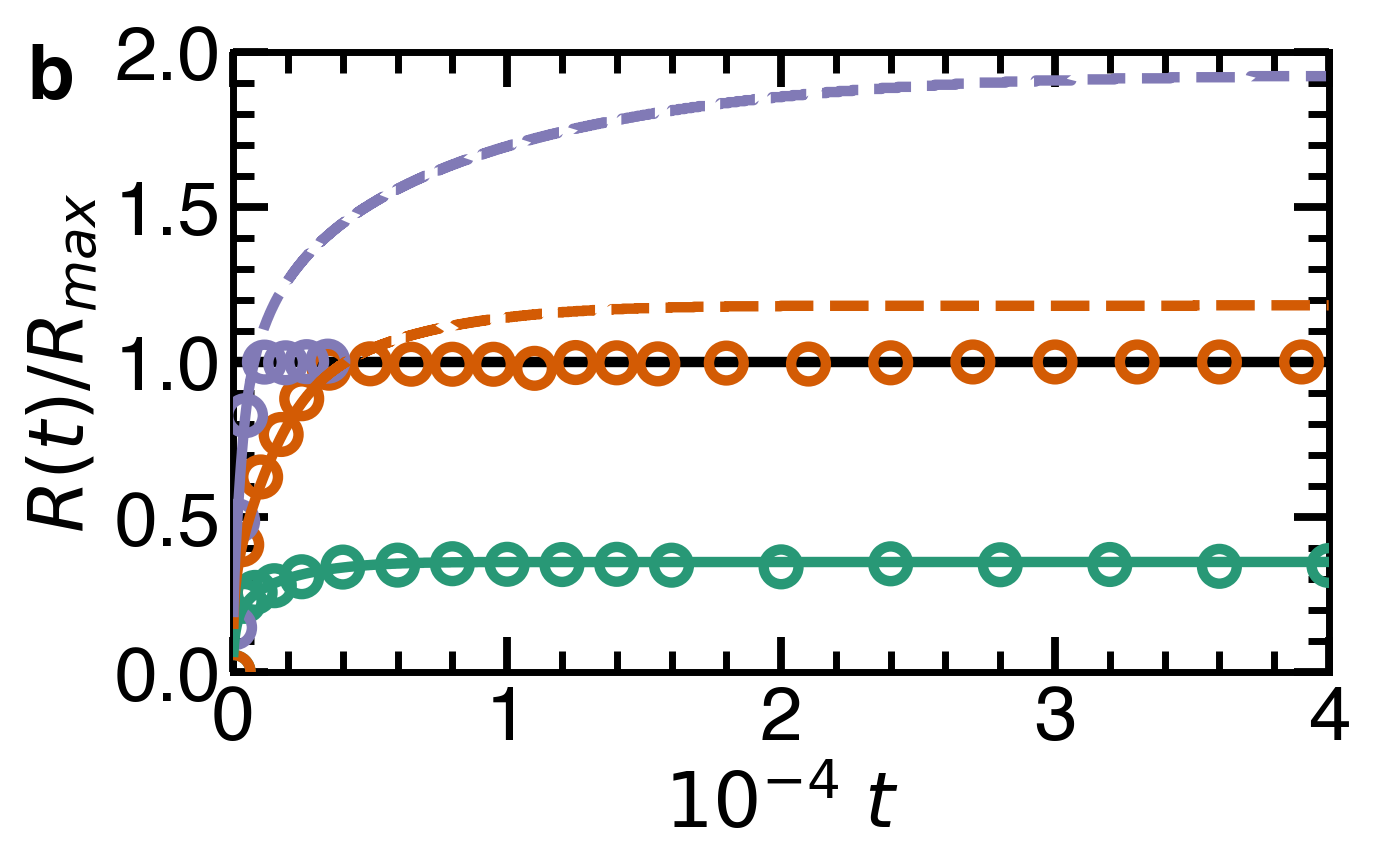}
\caption{The growth of the domain size $R(t)$ in the dissipative-force formulation of the chiral model a) for short growth times, and b) long growth times, for dihedral force constants {\it{$k_d$}}=5 (green), {\it{$k_d$}}=9.86 (orange) and {\it{$k_d$}}=19.86 (purple) at $P$=0.1 and {$T=0.8$}. $R(t)/R_{max}=1$ corresponds to the length scale of the simulation box. The size of a phase domain is restricted by this length scale, and thus, the dashed curves correspond to predictions of the domain growth that could be observed if it would not be restricted by the size of the simulation box. The open circles are computational data, while the solid and dashed curves are obtained from the maximum of the time-dependent structure factor for the domain growth - see Appendix {D} and reference\cite{MFT_PT_2021}. The steady-state limit of the size of a phase domain, $R_{\infty}$, is proportional to the magnitude of the dihedral force constant.}
\label{fgr:FigureThree}
\end{figure}

The {\it{$k_d$}} dependence of the temporal evolution of an equimolar mixture of enantiomorphs as the system moves from a homogeneous mixture towards phase separation is illustrated in Figures \ref*{fgr:FigureThree}a,b. In particular, these figures depict $R(t)$ normalized by the maximum phase separation length scale $\it{R_{max}}$ (i.e. when the domain growth is restricted by the finite size $\ell$ of the system) for short (Figure \ref*{fgr:FigureThree}a) and long (Figure \ref*{fgr:FigureThree}b) time scales. When $R(t)$ reaches $\it{R_{max}}$, the mixture is fully separated at the scale of the simulation box and the domains stop growing. The dashed curves indicate the theoretical prediction for the domain growth if it would not be restricted by the size of the simulation box {(see below and Appendix {D})}. At $k_d$=19.86, corresponding to a relatively rigid dihedral angle spring constant, the time of phase separation $\tau_{LLPS}$, defined as the time when $R(t)/\it{R_{max}}=1$, is about 800 (see Figure \ref*{fgr:FigureThree}a). When $\it{k_{d}}$=9.86, the time required for full phase separation increases to {\it{$\tau_{LLPS}$}}=4000.

\begin{figure}[htbp]
 \includegraphics[width=3.25in]{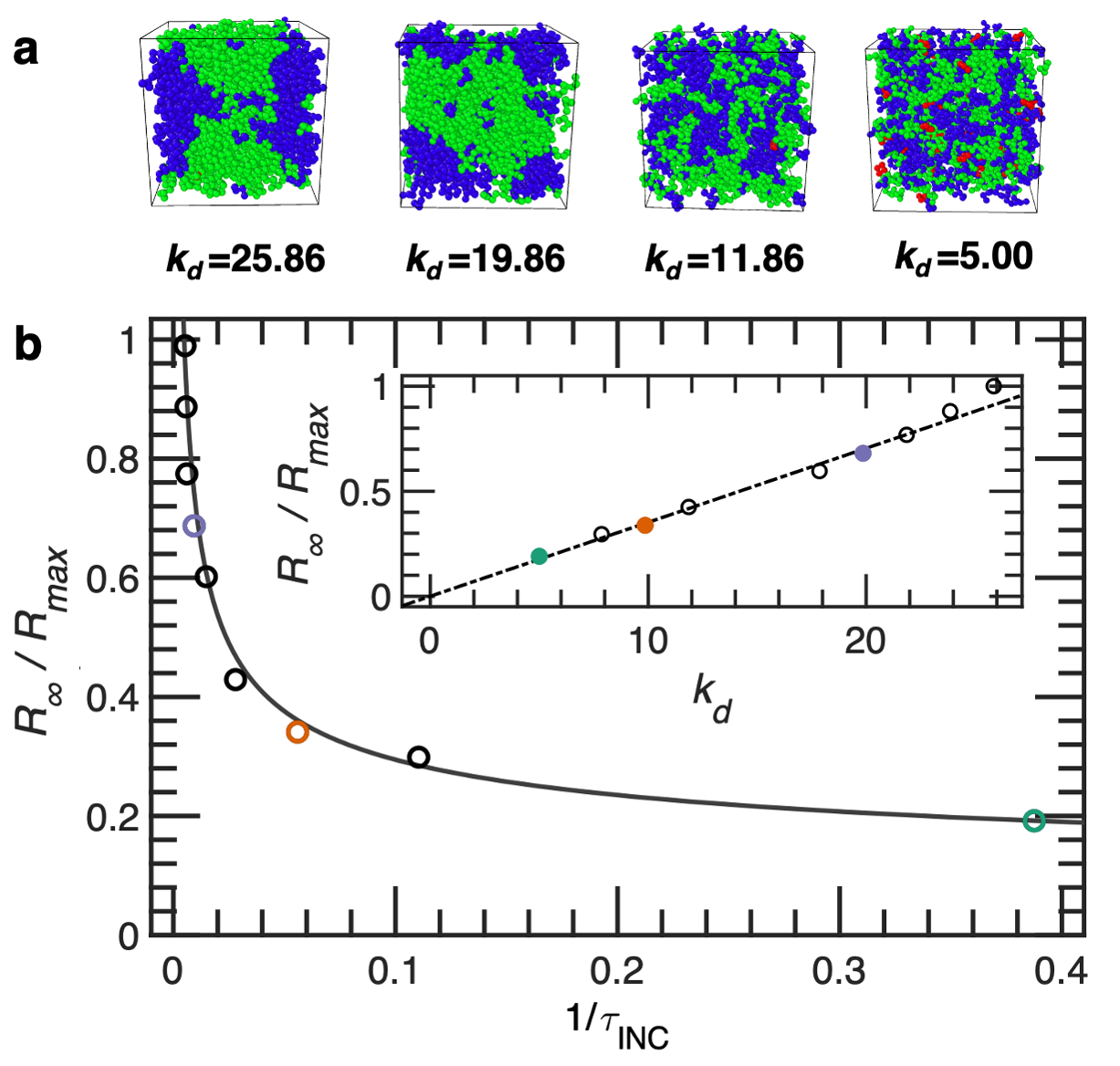}
\caption{The change of compositional heterogeneity with chiral interconversion kinetics at $T$=1.7 and $P$=0.1 in the dissipative-force formulation. a) Steady-state snapshots of chiral liquid systems at various dihedral force constants ({\it{$k_d$}}), b) The steady-state domain size as a function of interconversion rate, $1/\tau_{\text{INC}}$.  The solid line is the approximation given by $1/\tau_{INC} = a_1/R_{\infty}^2 + a_2/R_{\infty}^4$, where $a_1 = 4.6\times 10^{-3}$ and $a_2 = 3.8\times 10^{-4}$. In the first-order approximation, this follows from Equations \ref*{Eq-Interconversion_Coeff} \& \ref*{Eq-QCMinus}. The inset shows the linear correlation between $R_{\infty}$ and $k_d$. The colored points highlight the results corresponding to the three dihedral force constants for which the domain growth is shown in Figure \ref*{fgr:FigureThree}.}
\label{fgr:FigureFour}
\end{figure}

Figure \ref*{fgr:FigureThree}b illustrates the infinite time limit of the dependence of the steady-state domain length, $R(t{\to\infty})=\it{R_\infty} \propto k_d$, on the dihedral force constant of the tetramer model at $P$=0.1 when the system is quenched from $T$=2.6 to $T$=0.8. This dependence is consistent with the result presented in the inset in Figure \ref*{fgr:FigureFour}b, where $R_\infty$ is depicted as proportional to $k_d$. It is shown that below {\it{$k_d$}}=25.86, the domain growth saturates at a certain steady-state value below $\it{R_{max}}$, which is shown by the smaller tetramer inhomogeneities in Figure \ref*{fgr:FigureFour}a. The emergence of such smaller domain sizes suggests that the growth of the domains at these conditions is restricted by the dissipation in the intermolecular interaction forces, not by the finite length scale of the simulation box. {As the interconversion Onsager kinetic coefficient, $L${,} is inversely related to the characteristic time of interconversion, $\tau_{INC}$, we note that the data can be well described by $1/\tau_{INC} = a_1/R_{\infty}^2 + a_2/R_{\infty}^4$ (solid-line in Figure \ref*{fgr:FigureFour}b), consistent with the kinetics of the domain growth presented in Figure \ref*{fgr:FigureThree}b}. When {\it{$k_d$}}$=$25.86, the system reaches the onset of phase separation where the enantiomers completely phase separate at the size of the simulation box upon reaching equilibrium ($\it{R_{\infty}=R_{max}}$). Since $R_\infty \propto k_d$ and the interconversion rate is related to the kinetic Onsager coefficient as $L = 1/\tau_{INC}$, in which the term of order $1/R_\infty^4$ is negligible (for all $R_\infty > 1$) in the first order approximation, then $L$ is given as
\begin{equation}\label{Eq-Interconversion_Coeff}
    L = \frac{1}{\tau_{\text{INC}}} \approx M\frac{T^2}{k_d^2}
\end{equation}
where, {the squared temperature dependence comes from the natural coupling between dihedral angle rotation and thermal energy in an equilibrium ensemble. Furthermore, the assumption} $L \propto M$ {provides a good fit to simulation data (see Figure 8) and implies that enantiomer interconversion is linked to rotational mobility (the latter being proportional to translational mobility through the Stokes-Einstein and Debye-Stokes-Einstein equations.)}

\begin{figure}[h]
    \centering
    \includegraphics[width=\linewidth]{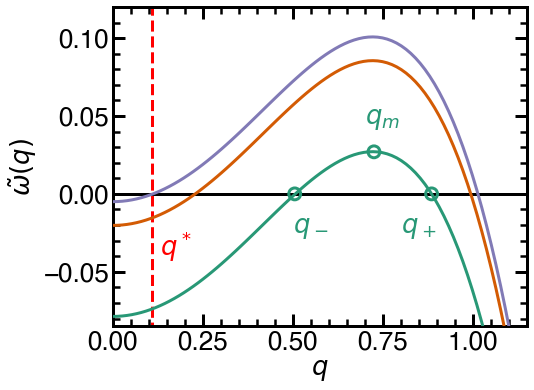}
    \caption{The growth rate, given by Equation \ref*{Eq-SourceAmplification}, at constant temperature for dihedral force constants: $k_d=5$ (green), $k_d=9.86$ (orange), and $k_d=19.86$ (purple); where $T=1.8$, {$T_\text{c}=2.3$}, $M = 0.8$, and $L$ is calculated from Equation \ref*{Eq-Interconversion_Coeff}. The red-dashed line corresponds to the inverse maximum size of the phase domain {$1/R_\text{max} \approx q^*\sim 1/\ell$. In this figure, we adopt $q^*=  0.11$} as obtained from the onset of phase separation on the length scale of the simulation box. The existence of a non-zero $q_- > q^*$ indicates the formation of steady-state microdomains.}
    \label{fig:GrowthRate}
\end{figure}

The generalized Cahn-Hilliard model with interconversion of species, introduced in the discussion of Equation \ref*{Eq-Amplification_Factor}, can be {adapted} to the dissipative chiral model by introducing an imbalance of chemical potentials, which produces nonequilibrium {forced} racemization (see Appendix B.2 and reference \cite{MFT_PT_2021} for details). {As a result,} the form of the growth rate for the dissipative formulation of the chiral model is 
\begin{equation}\label{Eq-SourceAmplification}
    \tilde{\omega}(q) = -L(\hat{T}+q^2) -M\Delta Tq^2(1-\xi^2q^2)
\end{equation}
where {$\hat{T}=T/T_\text{c}$} and $L$ is given by Equation \ref*{Eq-Interconversion_Coeff}. It is seen that the only difference between Equation \ref*{Eq-Amplification_Factor} and Equation \ref*{Eq-SourceAmplification} is that the interconversion Onsager kinetic coefficient, $L$, is decoupled from $\Delta T$. This equation is illustrated in Figure \ref*{fig:GrowthRate}, in which the effect of dissipation can be seen in the downward shift of the growth rate curve. We note that, in this form, this growth rate formula resembles the one introduced in Glotzer \textit{et al.}'s nonequilibrium lattice model, in which {forced} interconversion is decoupled from equilibrium phase separation \cite{GlotPhysRevLett.72.4109, GlotPhysRevLett.74.2034}. {{Glotzer and coworkers showed that phase separation driven by spinodal decomposition can be kinetically arrested at a certain scale due to the suppression of the growth of low wave-number inhomogeneities. Our computational results for the dissipative force formulation of the chiral model confirm the assessments made by Lefever \textit{et al}. \cite{P36hysRevLett.75.1674, P37hysRevE.56.3127}, and more recently, Lamorgeze and Mauri \cite{PLahysRevE.94.022605}, who argued that the results of Glotzer \textit{et al}. \cite{GlotPhysRevLett.72.4109, GlotPhysRevLett.74.2034} are limited to nonequilibrium conditions, where a source of forced interconversion, which inhibits the relaxation of the system to equilibrium, leads to the steady-state phenomenon of arrested phase separation into microdomains.}}

One can predict the size of the steady-state arrested domains from the condition that the growth rate becomes zero at a nonzero wave number, $q_-$,\cite{MFT_PT_2021} which is inversely proportional to the size of the domain, $q_- \propto 1/R_\infty$. Solving Equation \ref*{Eq-SourceAmplification} for this wave number {in the first order approximation ($T^2/k_d^2 \ll 1$)} gives
\begin{equation}\label{Eq-QCMinus}
    q_-^2 =\frac{L}{-D_\text{eff}}\approx \frac{T^2}{k_d^2 (-\Delta \hat{T})}
\end{equation}
{where  $D_\text{eff}=(M\Delta T+L)/\hat{T}$ is the effective mutual diffusion coefficient and $\Delta\hat{T}=\Delta T/\hat{T} = 1 - T_\text{c}/T$}. Indeed, as shown in the steady-state limit of Figure \ref*{fgr:FigureThree}b and in the inset of Figure \ref*{fgr:FigureFour}b, $R_{\infty}$ is proportional to $k_d$ as predicted by Equation \ref*{Eq-QCMinus}. {This equation illustrates the physics of microphase separation: the competition between racemization and negative diffusion.} 

Within our simulations, we observed a finite size effect, in which the small size of the simulation box limited the size of the steady-state microdomains, such that, computationally, it would appear as if complete phase separation was occurring. Such an effect may also be predicted from the growth rate formula and characteristic size, Equations (\ref*{Eq-SourceAmplification} \&\ref*{Eq-QCMinus}). For certain temperatures (at constant $k_d$), the characteristic wavelength, $q_-$, reaches the characteristic wavelength of the simulation box, $q^*$ (which is related to the size of the simulation box through $R_\text{max}\sim 1/q^*$), hence phase separation is observed on the length scale of the computation box. Since $q_-$ is cut-off at $q^*$, the temperature corresponding to the cut-off ($T^*$) characterizes the onset of the observed phase separation. The effect of this cut-off is consistent with what is illustrated in Figure \ref*{fgr:FigureThree}b, where (for $k_d=9.86$ and $k_d=19.86$) the computational data shows that the steady-state domain size stops growing when the system reaches the size of the simulation box.

\begin{figure}[h!]
\centering
\includegraphics[width=\linewidth]{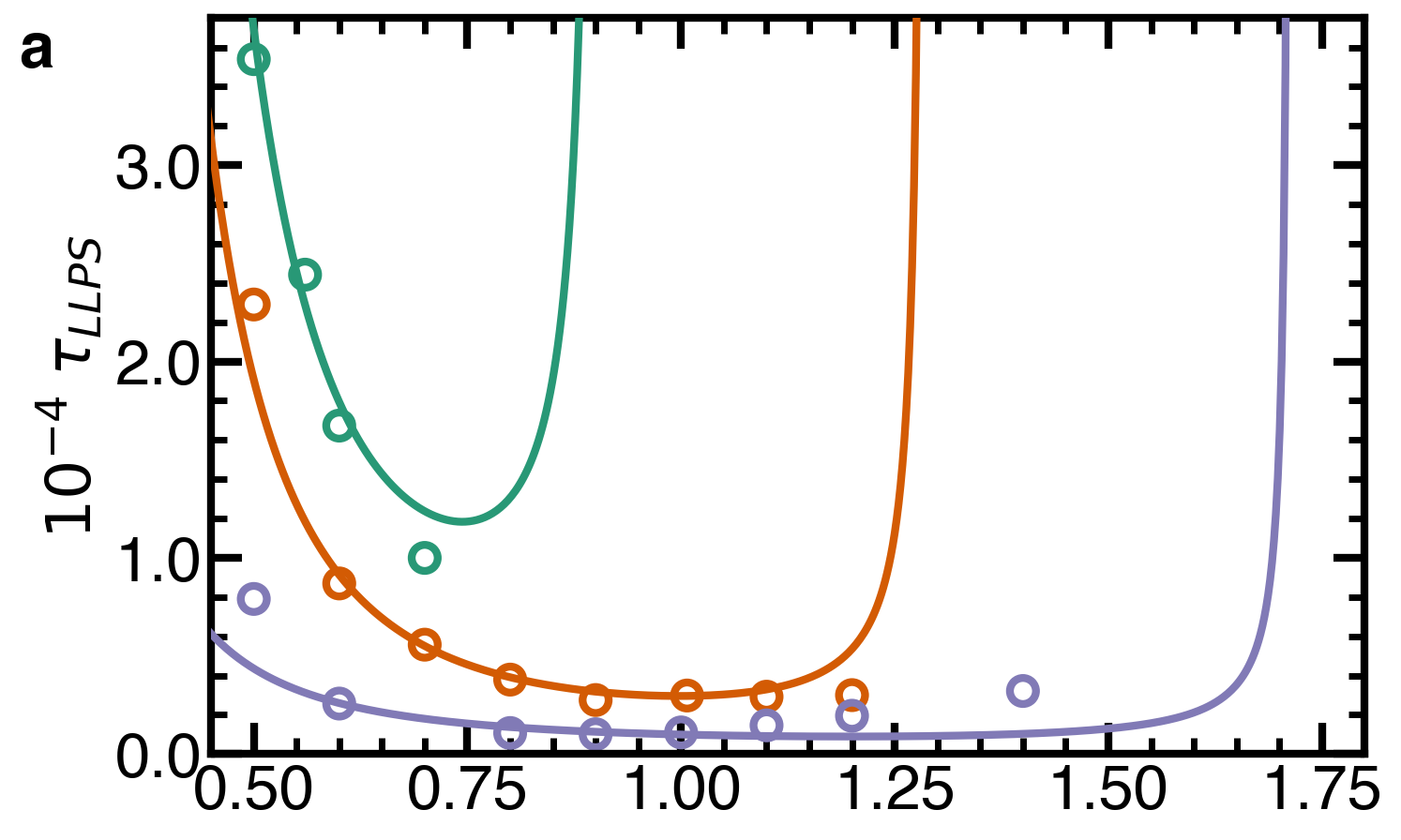}
\includegraphics[width=0.975\linewidth]{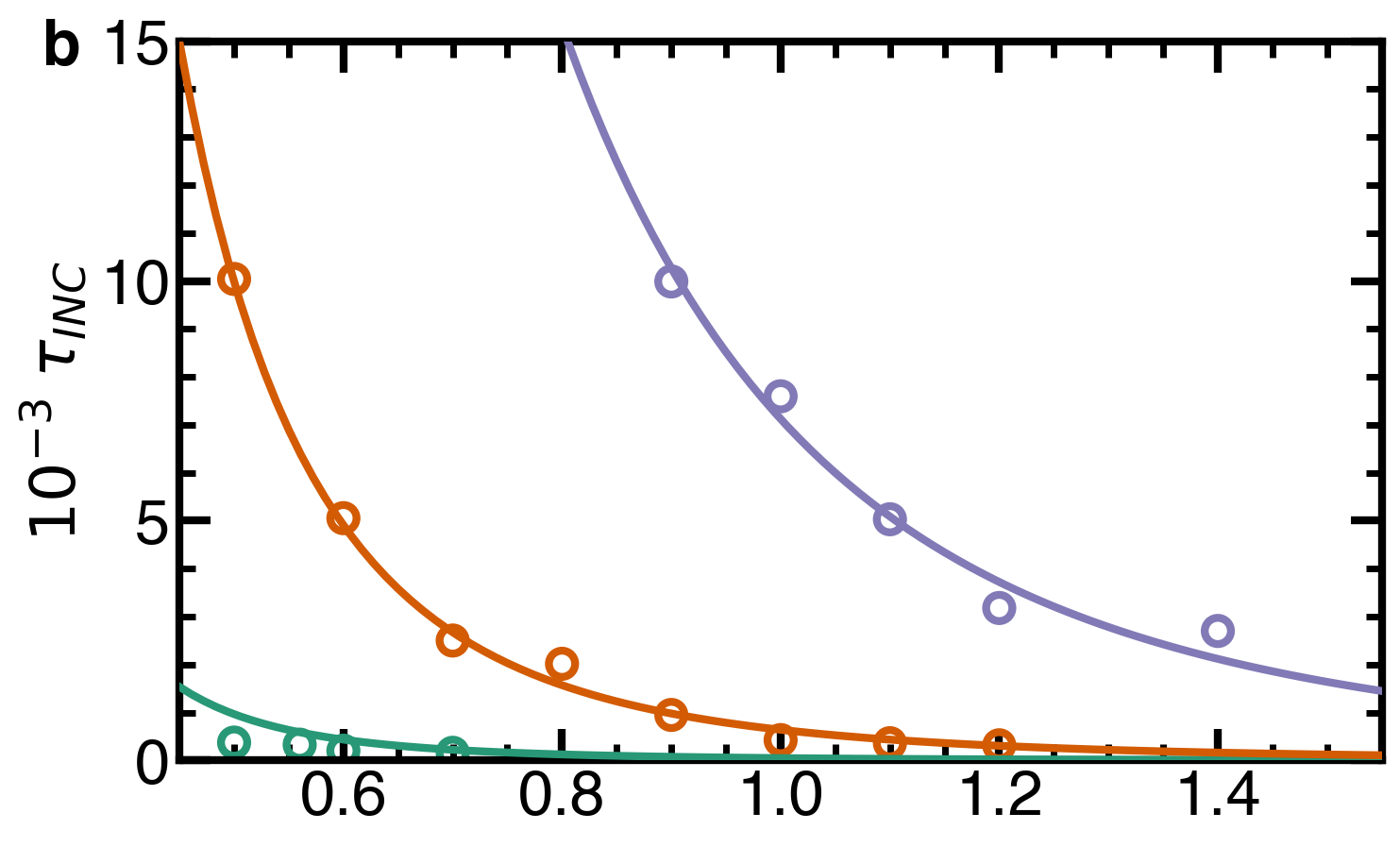}
\includegraphics[width=\linewidth]{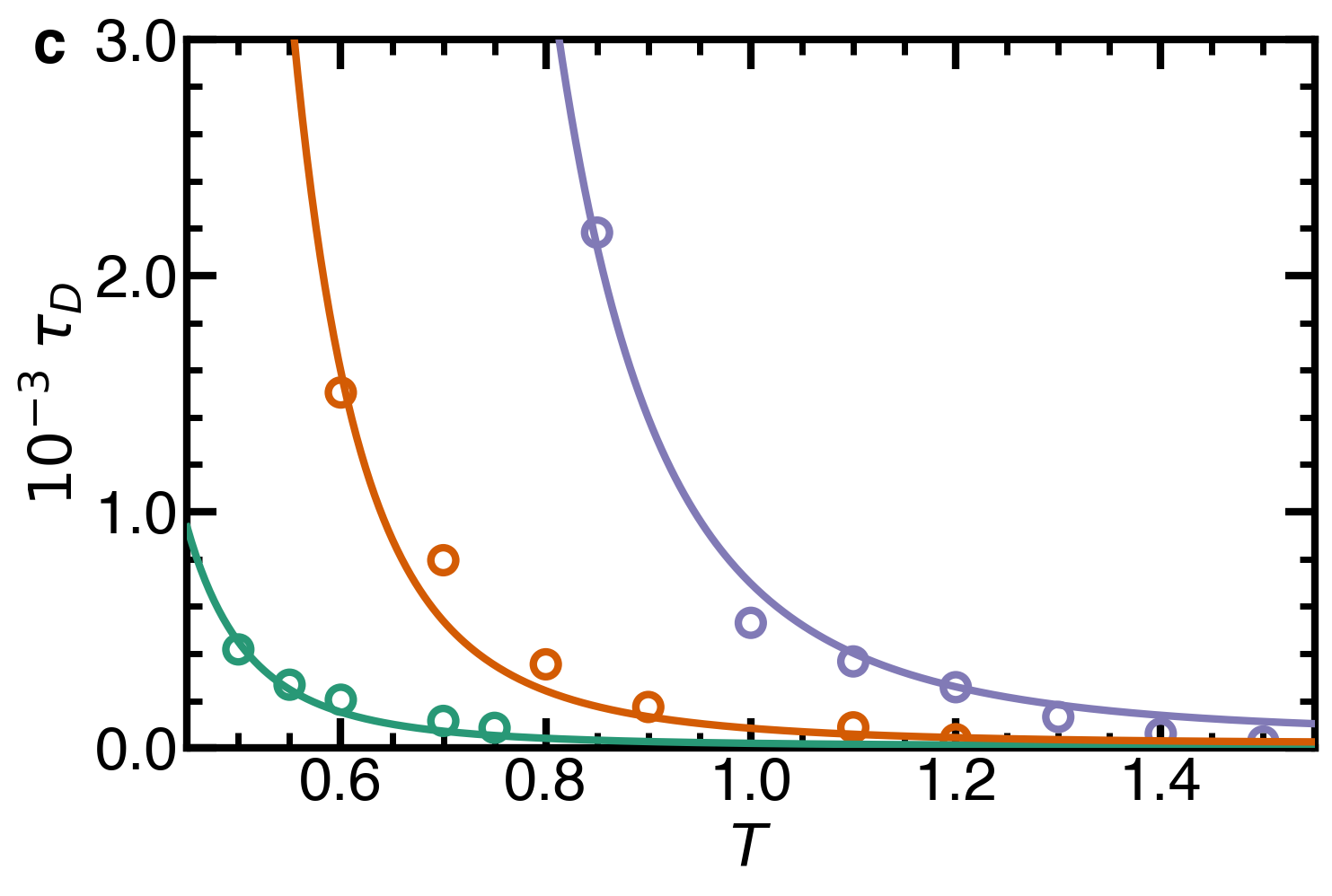}
\caption{Temperature dependence of the characteristic time scales in the dissipative-force formulation of the chiral model for dihedral force constants: $k_d=5$ (green), $k_d=9.86$ (orange), and $k_d=19.86$ (purple) at $P=0.1$. a) Characteristic time for liquid-liquid phase separation at the {length scale} of the simulation box, $q=q^*$. The curves are $\tau_{\text{LLPS}} \propto  1/\tilde{\omega}(q^*)$ where $\tilde{\omega}(q)$ is given by Equation \ref*{Eq-SourceAmplification} {where it was found that $q^* = 0.15$ and $T_\text{c} = 2.35$}. The condition $\tau_\text{LLPS}\to\infty$ corresponds to $T = T^*$.  b) Characteristic times for chiral interconversion. $\tau_{\text{INC}} \propto 1/L$, and the curves are given by Equation \ref*{Eqn_Lextended}. c) Characteristic self-diffusion times, and the curves are given by Equation \ref*{Eq_TauD2}. The parameters of Equations \ref*{Eq-SourceAmplification}, \ref*{Eqn_Lextended}, and \ref*{Eq_TauD2} used for the fits are given in Appendix {E}.}
\label{fgr:FigureFive}
\end{figure}

In order to further elucidate the kinetics of phase separation, we next consider the correlations between the characteristic interconversion time of a tetramer $\tau_{INC}$, the characteristic phase separation time $\tau_{LLPS}$, and the characteristic molecular self-diffusion time {\it{$\tau_D$}}, given by 
\begin{equation}\label{Eq_TauD1}
\tau_D = \frac{R_0^2}{D_\text{eff}}
\end{equation}
where $R_0$ is the radius of the tetramer's first solvation shell based on the site-site radial distribution function and $D_\text{eff}$ is the effective ($k_d$-dependent) self-diffusion coefficient computed from the slope of the time dependence of the mean-square displacement of all tetramers in the simulation box (Figure \ref*{fgr:FigureFive}). The circles {correspond to} temperatures below the onset of phase separation. Figure \ref*{fgr:FigureFive} demonstrates a strong coupling between the characteristic time of molecular diffusivity {\it{$\tau_{D}$}}, the characteristic time of phase separation {\it{$\tau_{LLPS}$}}, and the characteristic time of interconversion {\it{$\tau_{INC}$}}. 

In the two phase region, the growth rate formula, Equation \ref*{Eq-SourceAmplification}, describes the characteristic times of liquid-liquid phase separation (LLPS) on the length scale of the simulation box. As seen in Figure \ref*{fgr:FigureFive}a, the time of LLPS becomes infinite when the microdomain sizes reach the size of the simulation box, $q_- = q^*$. As indicated in this figure, this scenario corresponds to $T=T^*$. The time of LLPS affected by interconversion is defined through Equation \ref*{Eq-SourceAmplification} as $\tau_{\text{LLPS}} = a_\text{LLPS}/\tilde{\omega}(q^*)$, {where the amplitude $a_\text{LLPS} = 0.34$}, in the region where $q_- < q^*$ {(or, equivalently, where $R_{\infty}>\ell$)} as illustrated in Figure \ref*{fgr:FigureFive}a (see Appendix {E} for details). 

The interconversion time in the two phase region ($T < T^*$), as shown in Figure \ref*{fgr:FigureFive}b, is well-described by the extended version of Equation \ref*{Eq-Interconversion_Coeff} (Equation \ref*{Eqn_Lextended}) in which the next order, $k_d^4$ term, is included. In addition, the self-diffusion coefficient affected by interconversion is theoretically predicted from the coefficient of $q^2$ in $\tilde{\omega}(q)$ as given by Equation \ref*{Eq_SourceAmplification} (see also Appendix {E}), such that the effective diffusion coefficient, {$D_\text{eff} =  (M\Delta T + L)/\hat{T}$}, and Equation \ref*{Eq_TauD1} reads
\begin{equation}\label{Eq_TauD2}
    \tau_{D} = \frac{R_0^2}{(M\Delta T + L)/\hat{T}}
\end{equation}
The self-diffusion time shown in in Figure \ref*{fgr:FigureFive}c exhibits a crossover from the inverse mobility, {$\tau_{\text{D}} \propto \hat{T}/M\Delta T$ (at large $k_d$), to the intertconversion time, $\tau_{\text{D}} \propto \hat{T}/L$ (at small $k_d$).}

\begin{figure}[h]
\includegraphics[width=\linewidth]{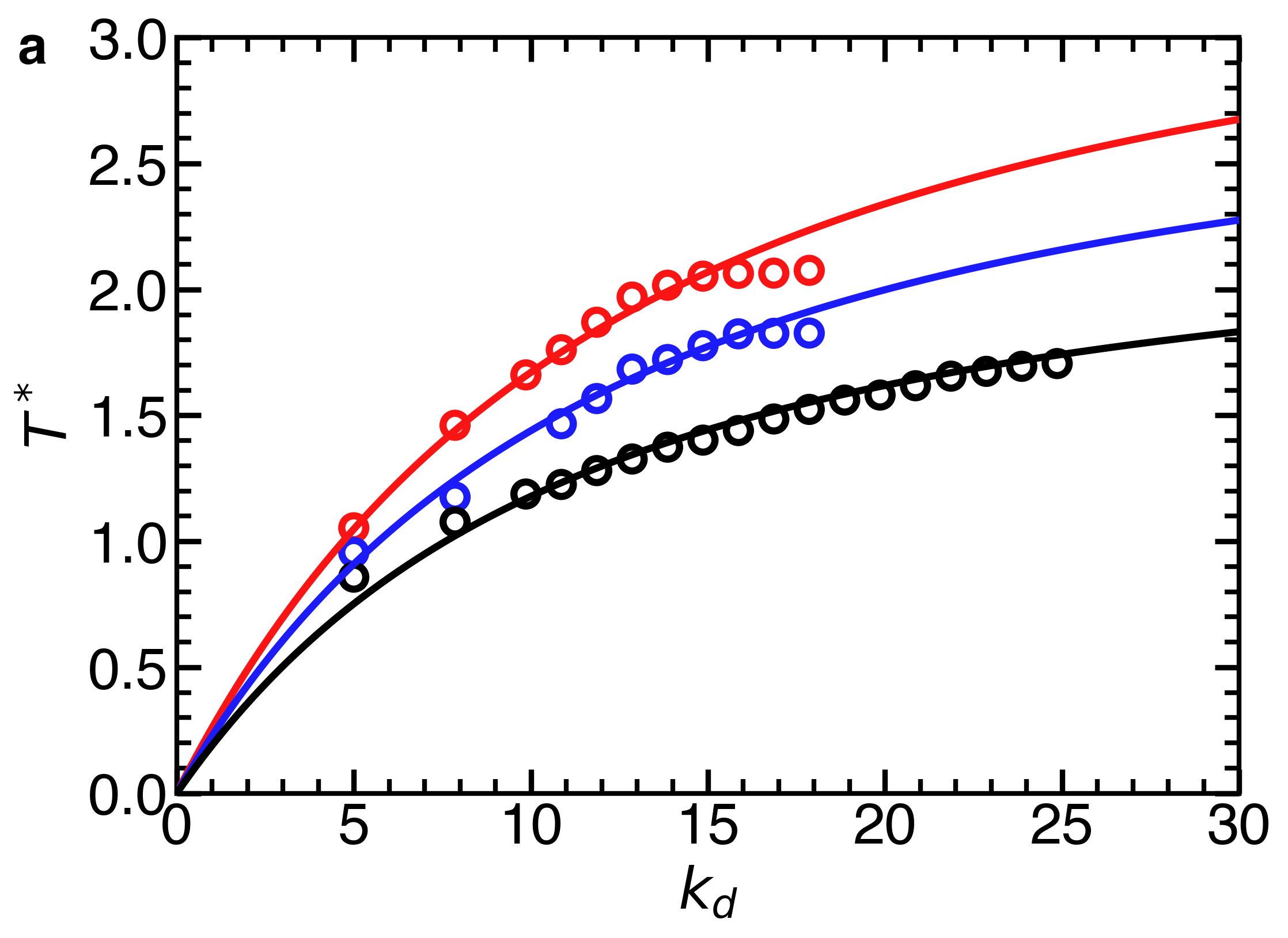}
\includegraphics[width=\linewidth]{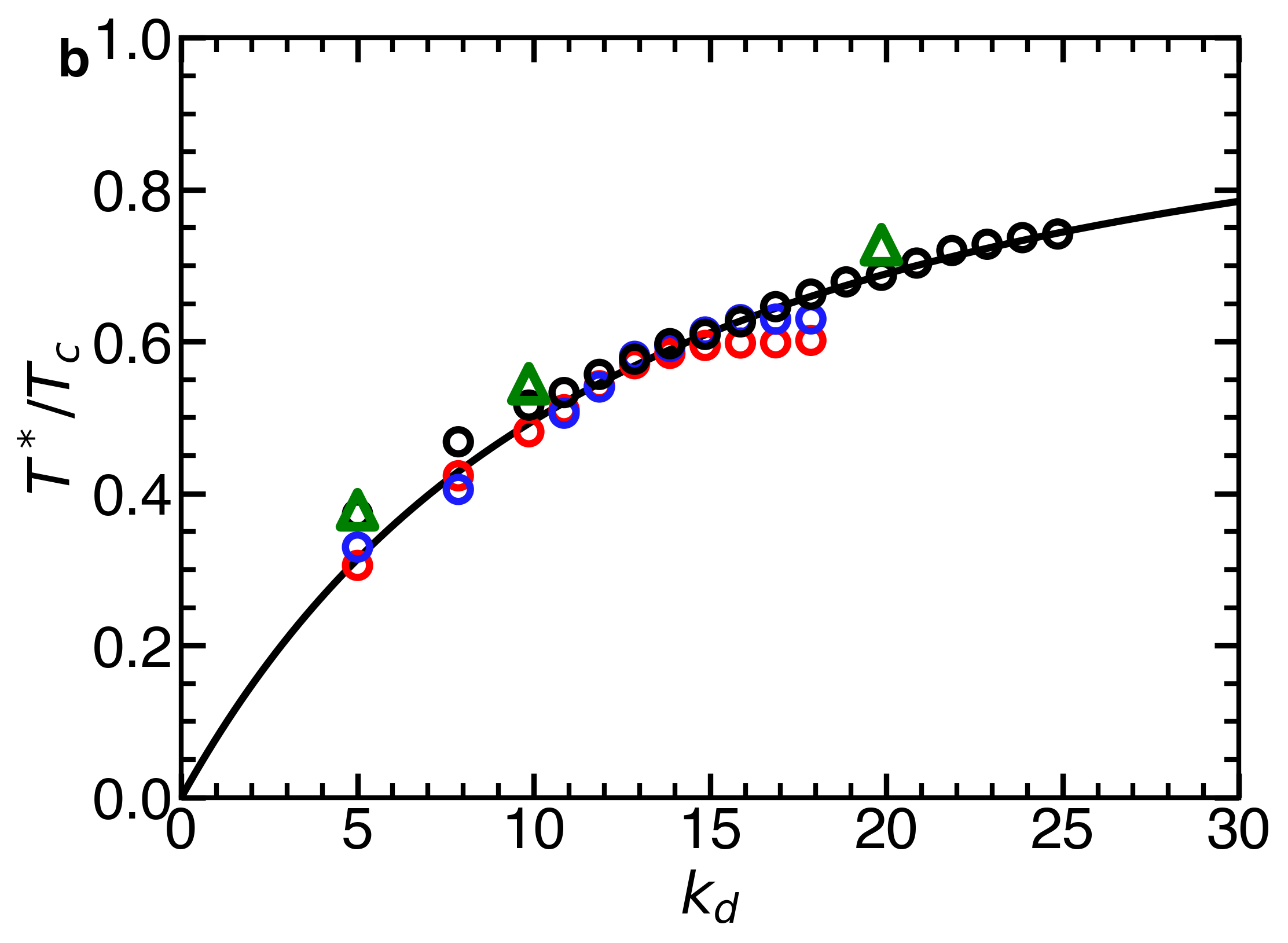}
\caption{Dihedral force constant dependence of the onset temperature of phase separation on the length scale of the simulation box in the dissipative-force formulation chiral model for $P=1.0$ (red circles), $P=0.5$ (blue circles), and $P=0.1$ (black circles). {The curves are numerically calculated from the first solution of $\tilde{\omega}=0$, given by Equation \ref*{Eq-SourceAmplification}, when $q_- = q^* = 0.11 \approx 1/R_\text{max}\sim 1/\ell$, $T = T^*$, and $T_\text{c}(P=1.0)=3.45$, $T_\text{c}(P=0.5)=2.91$, and $T_\text{c}(P=0.1)=2.3$} for different pressures (a) and in rescaled coordinates (b). The triangles, shown in (b), are obtained from the asymptotic limits of the time of liquid-liquid phase separation, $\tau_\text{LLPS}\to\infty$, as shown in Figure \ref*{fgr:FigureFive}a for $q^* = 0.15$ and $T_\text{c}=2.35$.}
\label{fgr:FigureSeven}
\end{figure}

{The asymptotic limit of $\tau_{\text{LLPS}}$ indicates the onset temperature, $T^*$, where phase separation occurs on the length scale of the simulation box. The onset temperature as a function of dihedral angle {force constant}, $k_d$, is depicted in Figure \ref*{fgr:FigureSeven}a. The onset temperature is numerically calculated from $\omega(q=q^*)=0$, given by  Equation 11. In the first order approximation, this solution is given by Equation 12 if $q_-=q^*$ and $T=T^*$}. As depicted in Figure \ref*{fgr:FigureSeven}a, the onset of arrested phase separation on the scale of the simulation box also depends on pressure. Physically, this originates from the density-dependent energetic bias towards homochiral interactions (Figure \ref*{fgr:FigureOne}), represented phenomenologically by the introduction of $\lambda > 0$ in the model. Thermodynamically, this pressure dependence of $T^*$ can be attributed to the underlying pressure dependence of the critical temperature of the liquid-liquid transition in the thermodynamic limit ($T_\text{c}$), at infinite $k_d$, due to the compressibility of the tetramer model. The values for the critical temperature for three different pressures are given in the caption of Figure \ref*{fgr:FigureSeven}a.

Rescaling the onset temperature by $T_\text{c}(P)$ gives the universal function of $k_d$ depicted in Figure \ref*{fgr:FigureSeven}b. We also show in Figure \ref*{fgr:FigureSeven}b, the predictions of $T^{*}$ obtained for three selected values of $k_d$ through the asymptotic limits of $\tau_{\text{LLPS}}\to\infty$ as illustrated in Figure \ref*{fgr:FigureFive}a. The predictions of $T^*$ are just above the observed onset temperature because they correspond to slightly higher values of $q^*$ and $T_\text{c}$ {(0.15 \textit{vs.} 0.11 and 2.35 \textit{vs.} 2.30 respectively). This difference can be attributed to uncertainty in obtaining the onset of phase separation on the scale of the simulation box.} These values, however, are in good agreement with the {computational} $T^*$ data obtained from the onset of phase separation.

As predicted from the growth rate factor, both conservative and dissipative force formulations will become identical in the limit of an infinitely rigid spring constant ($k_d\to\infty$) or when the kinetic interconversion Onsager coefficient goes to zero ($L\to 0$). We confirm this prediction by extrapolating the critical temperatures shown in Figure \ref*{fgr:FigureSeven}a to $k_d\to\infty$ and comparing $T_\text{c}(P)$ to the ones obtained from the conservative force formulation. Remarkably, this pressure dependence of the critical temperature for the conservative-force formulation of the chiral model is fully consistent with the prediction obtained from the dissipative-force formulation as shown in Figure \ref*{fgr:FigureTen}. This is evidence for the consistency of our computational data for these two alternative formulations of the chiral model.

\section{\label{sec:Conc}CONCLUSION}

The computational study of a three-dimensional {off-lattice} model of enantiomers with tunable chiral-interconversion kinetics reveals that arrested liquid-liquid phase separation into microdomains is observed when the intermolecular forces are not fully balanced, thus generating dissipation of energy which converts this model into a nonequilibrium steady-state model. This imbalance acts as a racemizing force that causes the arrest of the phase domain growth. In the conserved formulation, when the forces are balanced, the phenomenon of phase amplification, when one phase grows at the expense of the other, emerges, and the phase domain growth is only restricted by the system size. {From a numerical point of view, when} the dihedral force constant $k_d$ increases, the {kinetics of interconversion slow down correspondingly, causing phase amplification to slow down, and making it accordingly more difficult to observe the phenomenon on practical simulation times. In} the limit of $k_d\to\infty$ the system would undergo {the} usual phase separation\cite{Phase_Bullying_2021}.

{The physics driving amplification originates in the fact that molecules can interconvert, and ``species'' (in this case, molecules of type A and B) are not conserved. This provides the system with a mechanism for avoiding the energetically unfavorable formation of an interface between A-rich and B-rich phases, namely by committing (randomly, of course) to one or the other choice.}

This work can be extended to further investigate the role of the chiral bias parameter, $\lambda$, on interconversion and phase separation behavior. The tetramer model can also be generalized to consider nonzero enthalpy of interconversion where the interconversion rates for A$\rightarrow$B and B$\rightarrow$A could be different, thus the equilibrium concentration of the enantiomers would be a function of temperature. Generalization of the approach developed here for the particular case of an interconvertible chiral model could significantly improve fundamental understanding of the nature of phase behavior in a broad range of systems including polyamorphic liquids\cite{Anisimov_2018, Takae4471} and nonequilibrium phase separation of proteins into microdomains\cite{doi:10.1146/annurev-cellbio-100913-013325, Shineaaf4382, Wei_2017,ranganathan_liquid-liquid_2019}. {Examples of such possible generalizations include molecules whose local environment can fluctuate between (low-density, low-energy) and (high-density, high-energy) configurations (e.g., tetrahedral liquids, such as water or silicon), and proteins undergoing reversible structural fluctuations.}

\section{\label{sec:Ackn}Acknowledgements}
We thank Sergey V. Buldyrev for suggesting to consider {forced} interconversion through an imbalance of intermolecular forces and Nikolai D. Petsev for the development of the code for the conservative-force formulation of the chiral model. {M.A.A. and P.G.D. acknowledge the financial support of the National Science Foundation (awards CHE-1856479 and CHE-1856704, respectively.)} Simulations were performed on computational resources managed and supported by Princeton Research Computing, a consortium of groups including the Princeton Institute for Computational Science and Engineering (PICSciE) and the Office of Information Technology's High Performance Computing Center and Visualization Laboratory at Princeton University.
\nocite{*}
\bibliography{aipsamp.bib}

\appendix

\renewcommand\thefigure{\thesection.\arabic{figure}}   
\setcounter{figure}{0}  

\section{Molecular Simulation Details}

\begin{figure}[h!]
    \centering
    \includegraphics[width=\linewidth]{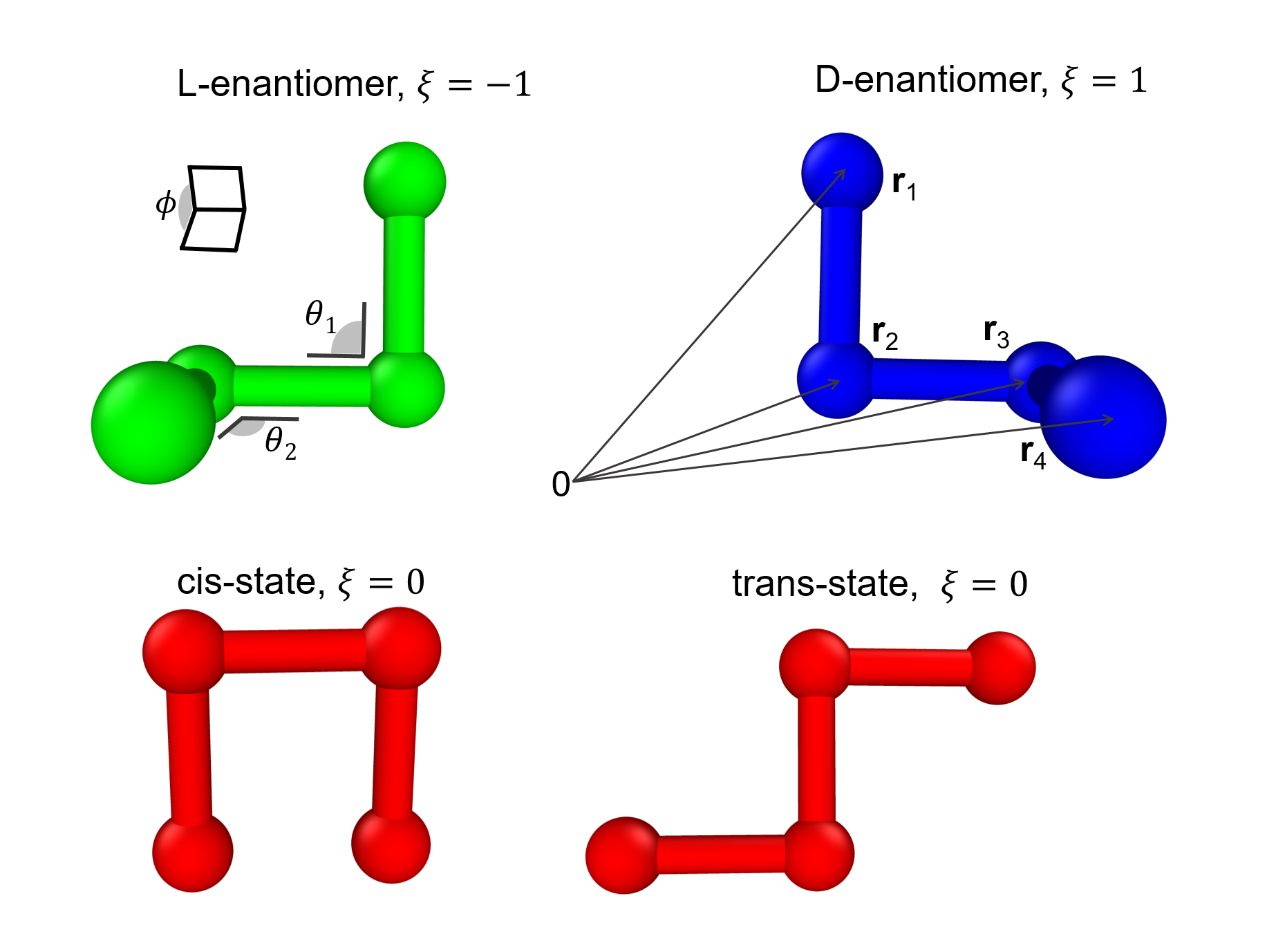}
    \caption{Molecular representation and relevant geometrical features of tetramer molecules. Molecules exist in left handed (A-type, green), right handed (B-type, blue) or achiral (cis- or trans-, red) configurations.}
    \label{Chirality_FigSupp1}
\end{figure}

We conducted the simulations of the models with energy conservation and energy dissipation using the molecular dynamics (MD) package, LAMMPS. MD simulations were performed
in an isothermal-isobaric ensemble. The temperature and pressure were controlled using Nosé-Hoover thermostats and barostats, respectively. Periodic boundary conditions were applied in three directions. The spherical cutoff for pair interactions was set to 3.5$\sigma_{tt}$ for the energy-conserving model and 3$\sigma_{tt}$ for the energy-dissipative model. A time step of 0.0005$t$ was used, in which $t$ is time in reduced units.

\setcounter{figure}{0}
\section{The Phase Domain Growth Rate}

\subsection{Conservative-Force Formulation of the Chiral Model}
When a binary mixture of two {non-interconverting} species is quenched from a high temperature to a low temperature, below the critical point of demixing, {and below the limit of compositional stability (spinodal curve),} then phase separation occurs through a process known as spinodal decomposition\cite{Cahn_1965}. By accounting for the interconversion {between} species present in the conservative-force formulation of the chiral model, then Cahn-Hilliard's theory of spinodal decomposition may be generalized to describe such a model. This is performed through the introduction of interconversion dynamics into the temporal evolution of the concentration of one of the enantiomer species, $c_A$, towards equilibrium. As a result, in the conservative-force formulation, the temporal evolution is described by 
\begin{equation}\label{Eq_TimeEvoEq}
    \frac{\partial \hat{c}_A}{\partial t} = M\nabla^2\mu -L\mu 
\end{equation}
where $M$ and $L$ are the diffusion and interconversion Onsager kinetic coefficients respectively, and $\mu$ is the chemical potential difference {between} the two alternative species $A$ and $B$ as, $\mu = \mu_A - \mu_B$, which is $\mu = 0$ in equilibrium. The reduced concentration of enantiomer species A, $\hat{c}_A$, is related to the physical concentration through $\hat{c}_A = 2(c_A - 1/2)$. The chemical potential for a binary mixture, which includes an additional contribution to the free energy due to the interfacial concentration profile, is determined through a Landau-Ginzburg free energy functional of the form
\begin{equation}\label{Eq_LGfunc}
      F({\hat{c}_A}) = \int\left({f_0(\hat{c}_A) +\frac{1}{2}R_0^2|\nabla \hat{c}_A|^2}\right)\text{d}{V}
\end{equation}
where the first term represents the thermodynamic ``bulk'' free energy {density} and the second term represents the contribution to the total free energy due to concentration inhomogeneities. In the second term, $R_0$ is the characteristic length scale of intermolecular interactions adopted as the size of a tetramer molecule ($R_0=1$ in reduced units). In the symmetric binary-lattice (``regular solution'') model formulated in the mean field approximation\cite{Anisimov_2018} $f_0$ can be expressed through the Gibbs energy of mixing, $\Delta G_\text{mix}$ as 
\begin{equation}\label{Eq_GibbsMix}
    f_0 = \frac{\Delta G_\text{mix}}{kT} = [c_A\ln{c_A} + (1-c_A)\ln{(1-c_A)}] + a c_A(1-c_A)
\end{equation}
where the term in brackets is the {ideal} entropy of mixing and the last term is the enthalpy of mixing. In the chiral model, $a \approx\rho \lambda$, where $\rho$ is the dimensionless density and $\lambda$ is chiral bias parameter. Expanding $f_0$ in the vicinity of the critical point, where $T_\text{c} = a/2$ and $\hat{c}_A =0$, in the lowest approximation $f_0 = (1/2)\Delta T\hat{c}_A^2$, where the reduced distance to the critical point $\Delta T = T/T_\text{c} - 1$. Therefore, the chemical potential is found from the variational derivative of the free energy functional with respect to the concentration of A-type enantiomers, as
\begin{equation}\label{Eq_GenMu}
    \mu = \frac{\delta F}{\delta c_A} = \frac{\partial f_0}{\partial \hat{c}_A}- \nabla^2 \hat{c}_A \approx \Delta T\hat{c}_A -\nabla^2 \hat{c}_A 
\end{equation}
Substituting Equation \ref*{Eq_GenMu} into Equation \ref*{Eq_TimeEvoEq}, the characteristic growth rate of the inhomogeneities, also referred to as the ``amplification factor'' \cite{Cahn_1965}, may be analytically determined with use of Fourier analysis to give
\begin{equation}\label{Eq_Amplification_Factor}
    \omega(q) = - \Delta T (L + Mq^2)(1-\xi^2 q^2)
\end{equation}
where $\xi$ is the correlation length of concentration fluctuations; in the mean field approximation, $\xi^2 \propto 1/(-\Delta T)$\cite{MFT_PT_2021}. In the absence of interconversion ($L=0$), Equation~\ref*{Eq_Amplification_Factor} reduces to the classical Cahn-Hilliard theory of spinodal decomposition\cite{Cahn_1965}.

\subsection{Dissipative-Force Formulation of the Chiral Model}
In the dissipative-force formulation of the chiral model, a sink of energy arises due to an imbalance in intermolecular forces, which {changes the chemical potential associated with the interconversion dynamics in the temporal evolution of the concentration, given by Equation \ref*{Eq_TimeEvoEq}}. Specifically, the imbalance in intermolecular forces alters the Gibbs energy of mixing, $\Delta G_\text{mix} = \Delta H_\text{mix} -T\Delta S_\text{mix}$, by effectively canceling the enthalpy of mixing\cite{MFT_PT_2021}. Therefore, in this formulation, the nonequilibrium free energy ($\tilde{f}_0$) is given by
\begin{equation}\label{Eq_GibbsMix_Noneq}
    \tilde{f}_0 = \frac{\Delta \tilde{G}_\text{mix}}{kT} = c_A\ln{c_A} + (1-c_A)\ln{(1-c_A)} 
\end{equation}
in which just the entropic contribution to the free energy remains. Thus, the energy dissipation forces the system into a ``racemized'' homogeneous state, which competes with the temporal evolution of the concentration towards thermodynamic equilibrium. Expanding the nonequilibrium free energy in the vicinity of the critical point to first order, we obtain {$\tilde{f}_0\approx -(1/2)\hat{T}\hat{c}_A^2$, where $\hat{T}=T/T_\text{c}$}. Therefore, since the imbalance in forces only affects the interconversion dynamics, then the temporal evolution of the concentration of species A, Equation \ref*{Eq_TimeEvoEq}, is modified to include a nonequilibrium chemical potential, {$\tilde{\mu}  \approx -\hat{T}\hat{c}_A -\nabla^2\hat{c}_A$}, and as such, is given by
\begin{equation}\label{Eq_TimeEvoNeq}
    \frac{\partial \hat{c}_A}{\partial t} = M\nabla^2\mu -L\tilde{\mu} 
\end{equation}
where the first term is the equilibrium diffusion dynamics, unaffected by the imbalance in forces, and the second term is the nonequilibrium interconversion dynamics. The Fourier analysis of Equation \ref*{Eq_TimeEvoNeq} gives the growth rate for the dissipative-force formulation of the chiral system in the form
\begin{equation}\label{Eq_SourceAmplification}
    \tilde{\omega}(q) = -L\hat{T} - \left(M\Delta T+L\right)q^2 - Mq^4
\end{equation}
which is an alternative form of Equation \ref*{Eq-SourceAmplification}, given in the main text, through the use of the mean field correlation length, $\xi^2 =-1/\Delta T$. \cite{MFT_PT_2021}

\setcounter{figure}{0}
\section{Dependence of Pair Interaction Energy on Dihedral Force Constant}

\begin{figure}[h!]
    \centering
    \includegraphics[width=\linewidth]{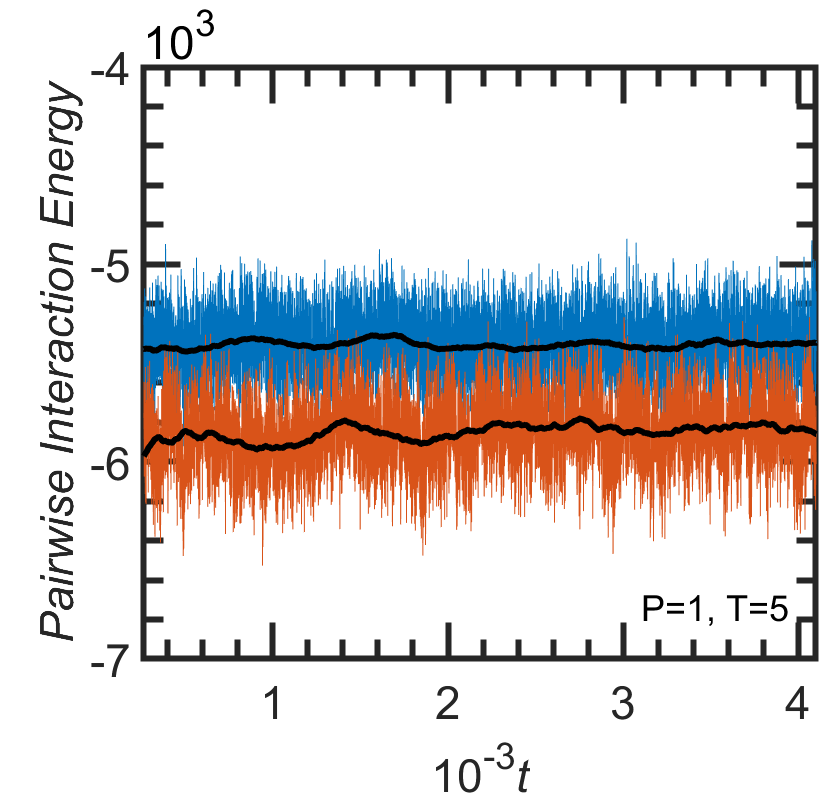} 
    \caption{Dependence of the total pair interaction energy on the dihedral force constant for the chiral model with conservative intermolecular forces at $k_d$ = 0.01 (blue) and $k_d$ = 30 (orange). The system consists of 1000 tetramers at $T$ = 5, $P$ =1 and $\left \langle EE \right \rangle$ = 0. The black lines are moving averages over time windows of duration $t = 200$.}
    \label{Figure_D1}
\end{figure}

\setcounter{figure}{0}
\section{Structure Factor and Domain Size Temporal Evolution}
The domain growth is determined through the time-dependent structure factor for the chiral system, $S(q,t)$. It is well-known that classical Cahn-Hilliard theory is only valid for the early stages of spinodal decomposition\cite{Cahn_Later_1966}. To accurately describe the behavior of systems in the late stages of spinodal decomposition, and eventually, the crossover into the coarsening regime, as reported by Langer \textit{et al}.\cite{langer_new_1975} and Binder \textit{et al}.\cite{Binder_Theory_1978}, two key alterations must be made to Cahn-Hilliard theory. First, concentration fluctuations, in the Ornstein-Zernike approximation\cite{Binder_Theory_1978} are introduced into the time-dependent structure factor\cite{cook_brownian_1970}, and second, the inverse susceptibility, $\partial^2f_0/ \partial\hat{c}_A^2$, must go to zero when the system reaches the spinodal. Adopting these changes into the time-dependent structure factor gives 
\begin{equation}\label{Eq_DiffStructFact}
    S(q,t) = \frac{Mq^2 + L}{-\tilde{\omega}(q)}\left(1 - e^{2\tilde{\omega}(q)t}\right)
\end{equation}
where $\Delta T$ in $\tilde{\omega}$, given by Equation \ref*{Eq_SourceAmplification}, becomes time dependent and is represented as
\begin{equation}
    \Delta T(t) = \Delta T(t=0) e^{-{t/\tau}}
\end{equation}
in which $\tau$ is a parameter that determines the transition from the early stages of spinodal decomposition to the coarsening regime.

\begin{figure}[h!]
    \centering
    \includegraphics[width=\linewidth]{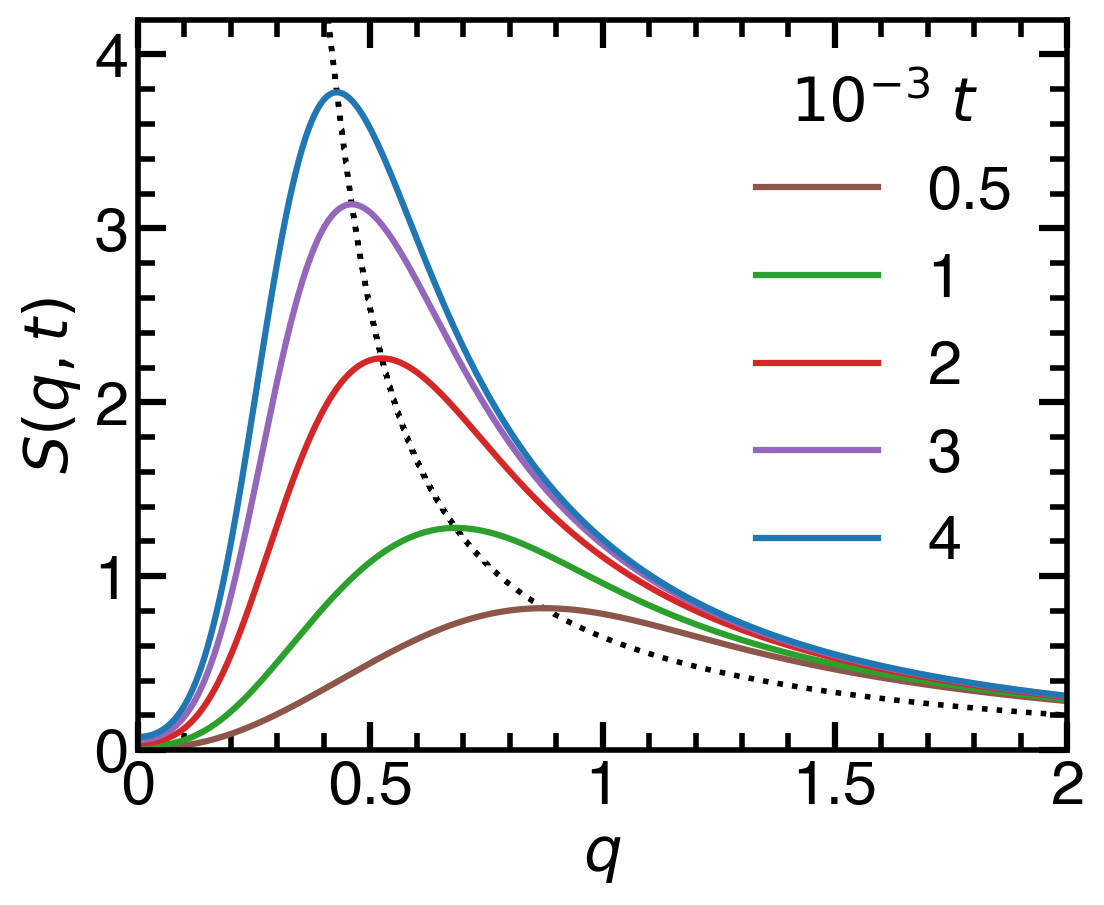}
    \caption{Structure factor as given by Equation \ref*{Eq_DiffStructFact} for $T=0.8$, $T_\text{c}=2.32$, $k_d = 9.86$, $P = 0.1$, $\tau = 1100$, $M$ given by the Einstein-Stokes relation, with $T_0 = 1.2$ and amplitude coefficient $b = 10^{-3}$, and $L$ given by Equation \ref*{Eq-Interconversion_Coeff}. Over time the maximum of $S(q,t)$, {indicated by the dotted line}, shifts to the left until {$q_m \propto q_-$},  but even in the $t\to\infty$ limit, the maximum never reaches zero wavenumber, thus corresponding to the formation of steady-state microdomains.}
    \label{SMFig_Sfactor}
\end{figure}

The domain size is determined from the characteristic wavenumber which corresponds to the maximum of the time-dependent structure factor. For instance, Figure \ref*{SMFig_Sfactor} shows the time evolution of the structure factor corresponding to $k_d = 9.86$ (whose time-dependent domain {size} is shown in Figure \ref*{fgr:FigureThree}). As illustrated, the maximum moves to the left until {$q_m \propto q_-$}, and if continued into the steady-state limit (when $t\to\infty$) the structure factor will never shift to zero wavenumber, indicating the formation of steady-state microdomains. The time evolution of the domain size (as presented in Figures \ref*{fgr:FigureThree}a,b) was determined from numerically calculating the time-dependent wave number corresponding to the maximum of the structure factor\cite{MFT_PT_2021}.

\setcounter{figure}{0}
\section{Characteristic Time Scales in the Dissipative-Force Formulation of the Chiral Model}

The characteristic time of liquid-liquid phase separation ($\tau_\text{LLPS}$), interconversion ($\tau_\text{INC}$), and molecular self-diffusion ($\tau_\text{D}$) are well described through the generalized Cahn-Hilliard theory. The interconversion Onsager kinetic coefficient, $L$, is given through an extended version of Equation \ref*{Eq-Interconversion_Coeff} (given in the main manuscript) of the form
\begin{equation}\label{Eqn_Lextended}
    L = \frac{1}{\tau_\text{INC}} = bM(T_0,T)\frac{T^2}{k_d^2}\left(1+c\frac{T^2}{k_d^2}\right)
\end{equation}
in which $b$ and $c$ are constants. The mobility, $M$, is given through the {Stokes-Einstein} relation that $M = kT/6\pi\eta R_0$, where the viscosity of the system is assumed to be $\eta = e^{T_0/T}$, in which the characteristic temperature, $T_0$, was also adjusted to better describe the behavior of each dihedral constant, $k_d$, at low temperatures. 

The characteristic time of liquid-liquid phase separation (LLPS) in the dissipative-force formulation of the chiral model can be determined from {the} region where $q_- < q^*$. In this region, the finite size of the system limits the size of the microdomains that may form, which is computationally observed as ``complete'' phase separation. Therefore, the characterisitc LLPS time is found as $\tau_\text{LLPS} = a_\text{LLPS}/\tilde{\omega}(q^*)$, where { the amplitude $a_\text{LLPS} = 0.34$ and $T_\text{c} = 2.35$}. In the fit, presented in Figure \ref*{fgr:FigureFive}a, $q^*$ is adjusted such that $\tau_{\text{LLPS}}\to\infty$ when the temperature reaches $T^*$, the onset of phase separation on the scale of the simulation box. Also, the characteristic temperature $T_0$ was slightly different for the three different dihedral constants: $T_0 = 2.2 (k_d = 5)$, $T_0 = 1.925 (k_d=9.86)$, and $T_0=1.2 (k_d=19.86)$.

The characteristic interconversion time, $\tau_{INC}$, of a tetramer is shown in Figure \ref*{fgr:FigureFive}b. It is described by Equation \ref*{Eqn_Lextended}, for which the constants were found to be $T_0 = 0.36$, $b = 29.4$, and $c = 526.6$. 

Equation \ref*{Eq_SourceAmplification} introduces an effective molecular self-diffusion coefficient, by the slope of the amplification factor at small wave numbers, which modifies the growth of the phase domain. This property is given by the coefficient of the $q^2$ term in Equation \ref*{Eq_SourceAmplification}, which reads as {$D_\text{eff} = (M\Delta T + L)/\hat{T}$}. As a result, the characteristic time of molecular self-diffusion (Figure \ref*{fgr:FigureFive}c) is found from, $\tau_D = a_D/D_\text{eff}$, where $a_D$ is an amplitude coefficient of the order $R_0^2$, {determined from the fit} to be $0.94$. Also, the characteristic temperature $T_0$ was found to be somewhat different for the three different dihedral constants: $T_0 = 3.4 (k_d = 5)$, $T_0 = 4.8 (k_d=9.86)$, and $T_0=7.0 (k_d=19.86)$.

\end{document}